\documentclass{emulateapj}
\usepackage{latexsym}
\usepackage{graphicx}
\shortauthors{West et al.}
\shorttitle{SDSS DR7 M Dwarfs}
\submitted{Accepted for Publication in AJ}
\begin{document}

\title{The Sloan Digital Sky Survey DR7 Spectroscopic M Dwarf Catalog I: Data}

\author{Andrew A. West\altaffilmark{1,2,10}, 
Dylan P. Morgan\altaffilmark{2},
John J. Bochanski\altaffilmark{3,9},
Jan Marie Andersen\altaffilmark{2},
Keaton J. Bell\altaffilmark{4},
Adam F. Kowalski\altaffilmark{4},
James R. A. Davenport\altaffilmark{4},
Suzanne L. Hawley\altaffilmark{4},
Sarah J. Schmidt\altaffilmark{4},
David Bernat\altaffilmark{5},
Eric J. Hilton\altaffilmark{4},
Philip Muirhead\altaffilmark{5},
Kevin R. Covey\altaffilmark{2,5,6},
B{\'a}rbara Rojas-Ayala\altaffilmark{5},
Everett Schlawin\altaffilmark{5},
Mary Gooding\altaffilmark{7},
Kyle Schluns\altaffilmark{2},
Saurav Dhital\altaffilmark{8},
J. Sebastian Pineda\altaffilmark{3},
David O. Jones\altaffilmark{2}}

\altaffiltext{1}{Corresponding author: aawest@bu.edu}
\altaffiltext{2}{Department of Astronomy, Boston University, 725 Commonwealth Ave, Boston, MA 02215}
\altaffiltext{3}{MIT Kavli Institute for Astrophysics and Space Research, 77
  Massachusetts Ave, Cambridge, MA 02139-4307}
\altaffiltext{4}{Department of Astronomy, University of Washington, Box 351580,
Seattle, WA 98195}
\altaffiltext{5}{Department of Astronomy, Cornell University, 610 Space Sciences Building, Ithaca, NY 14853}
\altaffiltext{6}{Hubble Fellow}
\altaffiltext{7}{Wells College, Department of Mathematical and Physical Sciences, 170 Main Street, Aurora, NY 13026 }
\altaffiltext{8}{Department of Physics and Astronomy, Vanderbilt University, 6301 Stevenson Center, Nashville, TN 37235}
\altaffiltext{9}{Astronomy and Astrophysics Department, The Pennsylvania State University, 525 Davey Lab, University Park, PA 16802}
\altaffiltext{10}{Visiting Investigator, Department of Terrestrial Magnetism, Carnegie Institute of Washington, 5241 Broad Branch Road, NW, Washington, DC 20015}

\begin{abstract} 
  We present a spectroscopic catalog of 70,841 visually inspected M
  dwarfs from the seventh data release of the Sloan Digital Sky Survey
  (SDSS).  For each spectrum, we provide measurements of the spectral
  type, a number of molecular bandheads, and the H$\alpha$, H$\beta$,
  H$\gamma$, H$\delta$ and Ca II K emission lines.  In addition, we
  calculate the metallicity-sensitive parameter $\zeta$ and identify a
  relationship between $\zeta$ and the $g-r$ and $r-z$ colors of M
  dwarfs.  We assess the precision of our spectral types (which were
  assigned by individual examination), review the bulk attributes of
  the sample, and examine the magnetic activity properties of M dwarfs, in
  particular those traced by the higher order Balmer transitions.  Our
  catalog is cross-matched to Two Micron All Sky Survey (2MASS)
  infrared data, and contains photometric distances for each star.
  Lastly, we identify eight new late-type M dwarfs that are possibly
  within 25 pc of the Sun.  Future studies will use these data to
  thoroughly examine magnetic activity and kinematics in late-type M
  dwarfs and examine the chemical and dynamical history of the local
  Milky Way.
\end{abstract}


\section{Introduction}

Over the past decade, wide-field, deep astronomical surveys have
provided an unprecedented statistical platform for studying the
Universe (e.g. Sloan Digital Sky Survey, Two Micron All Sky Survey,
2dF, UKIRT Infrared Deep Sky Survey).  For the lowest-mass and most
populous stars in the Milky Way (M dwarfs), the Sloan Digital Sky
Survey (SDSS) has yielded photometric samples that exceed 30 million
stars \citep[e.g.][]{boo10} and spectroscopic samples of over 40,000 M
and L dwarfs \citep{west08}.  Previous SDSS studies have elucidated
the mean properties of low-mass stars \citep{H02, west04, west05,
  bootem, schmidt_sam}, their magnetic activity and flaring properties
as a function of mass \citep{west04,west08, kowalski09,
  kruse10,hilton10}, as well as the structure and kinematics of the
local Milky Way thin and thick disks \citep{boomunn, juric08,
  fuchs09,boo10} and the low-mass initial mass and luminosity
functions \citep{covey08,boo10}.

M dwarfs have main sequence lifetimes that are considerably longer
than the age of the Universe and can be used to trace the evolution of
both stellar properties and the Milky Way disks.  \citet{west06,
  west08} showed that magnetic activity (as traced by H$\alpha$) in M
dwarfs decreases with age and that M dwarfs appear to have finite
activity lifetimes from $\sim$1-2 Gyr for early-type M dwarfs (M0-M3)
to $\sim$7-8 Gyr for later type stars (M5-M7).  These results are
important for the habitability of extrasolar planets orbiting M dwarfs
\citep[e.g.][]{mearth09}, as active stars may disrupt planetary
atmospheres \citep{segura10}.  The H$\alpha$ emission does not have
sufficient energy to significantly affect planetary systems but has
been found to be correlate with X-ray emission \citep{reid95,champ},
which can interact with extrasolar planet atmospheres.

While H$\alpha$ is the most commonly studied emission line in M
dwarfs, the higher-energy hydrogen Balmer and Ca II transitions are
also present in the optical spectra of active stars.  The higher
energy emission lines appear to trace different temperature regions in
the chromosphere and can be used to characterize the upper atmospheres
of active M dwarfs \citep[and references therein]{walkowicz09}.
However, lacking large samples of low-mass stars with spectroscopic
coverage across the entire optical bandpass, it is still not clear how
the various activity indicators trace each other in active M dwarfs
\citep{rm06,walkowicz09}.

\citet[hereafter W08]{west08} also showed that the ratio of CaH/TiO molecular indices
\citep[a quantity that is likely related to metallicity;][]{gizis97}
decreased with height in the Galactic disk (a proxy for age).  This
tantalizing result suggests that M dwarfs can be used to reconstruct
the chemical evolution of the local Milky Way disk.  However, the
inability to assign accurate metallicities to individual M dwarfs
limits a detailed exploration of Galactic chemical evolution.  Several
recent studies have made progress in calibrating the M dwarf
metallicity scale.  \citet{babs10} used infrared spectroscopy of M
dwarfs in wide binaries with higher-mass stars to calibrate infrared
spectral features with metallicity.  In the optical, \citet{lepine07}
developed a metallicity-dependent quantity $\zeta$, which is a relation
between the CaH and TiO molecular indices and defined as:

\begin{equation}
\zeta=\frac{1-\rm{TiO5}}{1-[\rm{TiO5}]_{Z_{\odot}}},
\end{equation}

\noindent where

\begin{eqnarray}
[\rm{TiO5}]_{Z_{\odot}}=& -0.164(\rm{CaH2+CaH3})^3 \nonumber\\
 & + 0.670(\rm{CaH2+CaH3})^2 \\
& -0.118(\rm{CaH2+CaH3})- 0.050.\nonumber
\end{eqnarray}

\noindent Using wide binaries consisting of an M dwarf and a higher
mass star, \citet{woolf09} showed that the $\zeta$ index is a good
discriminant between high and low-metallicity ([Fe/H] =0 and [Fe/H] =
-1 respectively) for early-type M dwarfs ($\sim$M0-M3).  While $\zeta$ is 
a useful tool for examining the relative metallicity among
stars, finer determinations of an absolute metallicity from
$\zeta$ will require additional calibration.

Before the SDSS \citep{york00}, the largest spectroscopic samples of M
dwarfs contained only a few thousand individual stars \citep{pmsu1}
and were primarily focused on the red portion of the optical spectrum
($\sim$6000-8000 \AA).  The spectroscopic catalogs of M dwarfs from
the first few SDSS data releases quickly surpassed previous records
for sample size \citep{H02, west04} and had spectral coverage that
spanned the entire optical band ($\sim$3900-9300 \AA) at low resolution (R$\sim$1800). 
\nocite{west08}W08 presented a spectroscopic catalog of more than
44,000 M dwarfs from the SDSS Data Release 5 (DR5), more than doubling
the M dwarf spectroscopic tally.  Due to the large number of spectra
in the DR5 sample, spectral types were assigned using the
Hammer\footnote{The Hammer software is available for public download
  at: http://astro.washington.edu/users/slh/hammer }automatic spectral
typing facility \citep{covey07}.  While the Hammer spectral types are
generally good to within $\pm$1 sub--type, there have been recent
indications that there may be systematic offsets for some of the M
dwarf spectral types (see Section 2).  Therefore, to ensure the
highest quality sample, it is important to visually inspect each
spectrum.  There have been two subsequent SDSS data releases since the
\nocite{west08}W08 sample that include over 50,000 additional M dwarf
candidates and contain many sightlines at low Galactic latitudes
\citep[as part of the SEGUE survey;][]{segue}.  The new lines of sight
extend both the radial and vertical extent of the M dwarf sample, and
provide a larger statistical platform from which to probe the
structure, kinematics and evolution of the Milky Way using its
smallest stellar constituents.

Here we present the latest spectroscopic catalog of M dwarfs
from Data Release 7 \citep[DR7;][]{dr7} of the SDSS.  While the main
thrust of this paper is the presentation and characterization of a new
spectroscopic sample, we have included a general magnetic activity
analysis as well as a discussion about the future use of this sample
to probe Galactic chemical evolution.  We describe our sample
selection techniques in Section 2, specifically discussing our manual
inspection of \emph{all} M dwarf candidates.  Section 3 gives the
mean properties of the sample, including a new activity analysis of
the hydrogen Balmer and Ca II K emission features.  Section 3 also
contains a list of candidate M dwarfs within 25 pc and explores how
the metallicity sensitive parameter $\zeta$ varies as a function of
stellar parameters.  We discuss our results and future use of our new
sample in Section 4.

\section{Data and Sample Selection}

The SDSS provides a large, uniform photometric and spectroscopic
dataset from which to extract high-quality samples of low-mass dwarfs
\citep{gunn98,fukugita96, hogg01, gunn06, ivezic04,pier03,
  smith02,tucker06}.  We used the DR7
CasJobs\footnote{http://casjobs.sdss.org/} tool to select 116,161 M
dwarf candidate objects with SDSS spectra that 1) fell within the
typical color range for M and L dwarfs \citep[$r-i > 0.42$ and $i-z >
0.24$; W08,][]{kowalski09}; 2) had SDSS spectral classifications of
``STAR'' or ``STAR\_LATE''; and 3) had radial velocities smaller than
1500 km\ s$^{-1}$.  \citet{kowalski09} found that the color cuts used
in previous studies did not extend blue enough ($r-i$ in particular)
to include all of the M0 dwarfs.  Our new (bluer) color selection corrects
for this previous oversight.

The spectra for all 116,161 candidates were visually inspected
and assigned spectral types using the Hammer spectral typing facility
\citep{covey07}.  Low signal-to-noise ratio (SNR) spectra (SNR $<$3 at
$\sim$8300 \AA) and extragalactic interlopers were removed during
visual inspection, resulting in 109,639 objects.  We also removed stellar spectra
that were not identified as M dwarfs during visual inspection as well as those spectra that
were duplicated in our sample (including 6771 spectra for 2661 M
dwarfs that were taken on different days) and arrived at the 70,841
spectra of M dwarfs in our DR7 catalog.  While many of the
objects that we removed were K or L dwarfs, others were white dwarf--M
dwarf (WD-dM) pairs and low-metallicity subdwarfs.  All of the L
dwarfs that were removed from our DR7 sample were cataloged and
analyzed by \citet{schmidt_sam}. Many of the other objects that were
not included in our DR7 sample will be presented in future
studies. Not all of the WD-dM pairs were removed during the visual
inspection.  We used the color cuts from \citet[][$u-g$ $<$ 2, $g-r$
$>$ 0.3, $r-i$ $>$ 0.7, $\sigma_{u,g,r,i}<0.1$]{Smolcic04} to identify
additional WD-dM pairs (that avoided manual detection).  All of the
497 objects that matched these color criteria are flagged as ``WDM''
in our DR7 catalog.  We also defined a clean photometric sample using
the SDSS photometric processing flags (SATURATED, PEAKCENTER,
NOTCHECKED, PSF\_FLUX\_INTERP, INTERP\_CENTER, BAD\_COUNTS\_ERROR were
all set to zero in the $r$, $i$, and $z$
bands)\footnote{http://www.sdss.org/dr7/products/catalogs/flags.html},
which resulted in 65,277 M dwarfs.  While our catalog contains all of the
70,84 visually inspected M dwarf spectra, the ``GOODPHOT'' and ``WDM'' flags
can be used to obtain samples that include good photometry and remove
possible WD-dM pairs respectively.

Radial velocities (RVs) were measured by cross-correlating each
spectrum with the appropriate \citet{bootem} M dwarf template.  This
method has been shown to produce uncertainties ranging from 7-10 km\
s$^{-1}$ \citep{bootem}.  All of the DR7 objects were cross-matched to
the USNO-B/SDSS proper motion catalog \citep{munn04,munn08},
identifying 39,151 M dwarfs with good proper motions\footnote{Good
  proper motions are defined as those with MATCH =1, DIST22 $>$ 7,
  SIGRA $<$ 1000, SIGDEC $<$ 1000 and (NFIT = 6 or (NFIT = 5 and (O
  $<$ 2 or J $<$ 2))) \citep{munn04}.}.  Distances to each star were
calculated using the $M_r$ vs.\ $r-z$ color-magnitude relation given
in \citet{boo10}.  Our calculated distances have uncertainties of
$\sim$20\%, arising mostly from the intrinsic spread of the main
sequence.  The proper motions and distances were combined with the RVs
to produce 3-dimensional space motions for the DR7 M dwarfs.  Although
we include the standard $U$, $V$, $W$ space motions in our catalog, we
caution that the $U$, $V$, $W$ velocities are in a Cartesian
coordinate system that may not be appropriate for stars at appreciable
distances from the Sun.  We therefore also include the Galactic radial
($R$), tangential ($\Theta$) and vertical ($Z$) cylindrical components
of the position and velocity for each star for which we have 3D space
motions.

We also matched our catalog to the 2MASS point source catalog
\citep{2mass}, matching only to unique 2MASS counterparts within
5$^{\prime\prime}$ of the SDSS position that do not fall within the
boundaries of an extended source ({\sc gal\_contam} $=$0).  To ensure
that we used only high quality 2MASS photometric data, we applied
additional cuts to each of the $J$, $H$, and $K_S$ bands. If the
source was not detected ({\sc rd\_flg} $=$0), nominally detected ({\sc
  rd\_flg} $=$6), was detected but unresolved ({\sc rd\_flg} $=$9), or
had contaminated/confused photometry ({\sc cc\_flag} $\neq$0) in a
particular band, the 2MASS data were not included.  This resulted in
57,956 2MASS counterparts with $J$, $H$ and $K_S$ magnitudes and their
uncertainties that were included in the catalog. 

We used the 2MASS photometry to investigate any possible M giant
contamination.  \citet{bessell88} found that M giants and M dwarfs
separate in $J-H$ vs. $H-K$ color space due to differences in H$_2$O
absorption in their atmospheres. By comparing the DR7 $J-H$ vs. $H-K$
color-color diagram with that of \citet{bessell88}, we find that no
more than 0.5\% of our sample could be giants.  In addition,
\citet{covey08} conducted a complete magnitude-limited spectroscopic
survey of a 1$\Box^{\circ}$ field in the SDSS footprint and concluded
that the giant contamination rate was less than 2\% for stars redder
than a spectral type of K5.

\subsection{Spectral Typing}

\begin{figure}
\plotone{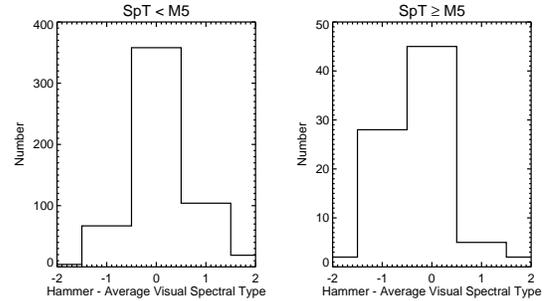}
\caption{The distribution of differences between the Hammer automatic
  spectral types and the mean of the visually inspected spectral types
  for M dwarfs with visual types earlier than M5 (left) and M5 and
  later (right).  While there is good agreement for the early-type M
  dwarfs, there is a systematic offset in the automatic types assigned
  by the Hammer; the Hammer assigns an earlier type $\sim$38\% of the
  time.}
\label{hamdiff}
\end{figure}

Previous large spectroscopic samples of M dwarfs have relied on
automated spectral typing due to the quantity of time required to
manually inspect tens of thousands of candidate spectra. Systematic
discrepancies were recently identified in the spectral types from the
\nocite{west08}W08 SDSS sample that were determined automatically by the
``Hammer'' spectral typing algorithm \citep{covey07}.  The bias was
detected as a systematic offset for late-type stars, whose automatic
classification was often 1 subtype earlier than determined via visual
inspection (see Figure \ref{hamdiff}).  We thus decided that visual
inspection would produce the most reliable and precise spectroscopic
sample.  We visually inspected all 116,161 M dwarf candidates and
manually assigned spectral types.  The sample was divided among 17
individuals\footnote{The order of the co-authors was based on the
  number of spectra examined.} who used the manual ``eyecheck'' mode
of the Hammer (v.\ 1\_2\_5) to assign spectral types and remove non-M
dwarf interlopers.

Figure \ref{hamdiff} shows the difference between the spectral types
automatically determined by the Hammer and the mean visual inspection.
For early-type M dwarfs (left panel), the Hammer and the spectral
typers agree most of the time.  However, for $\sim$38\% of the
late-type M dwarfs (right panel), the Hammer assigns spectral types 1
subtype earlier than the average human spectral typer.  This confirms
that while the Hammer generates automatic spectral types within the
quoted $\pm1$ subtype accuracy \citep{covey07}, there is a systematic
offset of 1 subtype for the late-type stars, and justifies our effort
to manually inspect more than 116,000 candidates.

Each spectral typer also examined a control sample of 1000 spectra, of
which 638 were M dwarfs.  We used this control sample to assess the
quality and reliability of the visually inspected sample by
quantifying the variations among the 17 individual typers, and by
comparing the median visual type for each star to its automatic Hammer
type.  The results of the control sample show that there is excellent
agreement between all of the spectral typers with a large fraction of
the stars being assigned.  Almost all of the visual classifications in
the control sample agreed to within $\pm$1 subtype of the median
value; for most of the stars, the dispersion in visual classifications
was $<0.4$ subtypes. To ensure that there were no major systematics in
the spectral types assigned by any individual, we examined the mean
difference between each typer's visual classifications of the control
sample spectra and the median classifications of the entire group.  We
divided the sample into the early types ($<$ M5) and late types
($\geq$ M5) to investigate any spectral type dependence in typing
quality: all of the spectral typers clustered near the median values
with no individual displaying a systematic offset larger than 0.2
subtypes.



\subsection{Measured Quantities}

As part of our analysis we measured a number of spectral lines and
molecular features in each M dwarf spectrum.  All of the spectral
measurements were made using the RV corrected spectra.  The TiO1,
TiO2, TiO3, TiO4, TiO5, TiO8, CaOH, CaH1, CaH2, and CaH3 molecular
bandhead indices and their formal uncertainties were measured using
the Hammer with the molecular bandheads as defined by \citet{pmsu1}
and \citet{gizis97}.  We also measured the chromospheric hydrogen
Balmer and Ca II lines that are associated with magnetic activity.  We
expanded the H$\alpha$ analysis of \citet{west04, west08} to include
H$\beta$, H$\gamma$, H$\delta$ and Ca II K (H$\epsilon$ and Ca II H
are blended in SDSS data and were not included in our sample).  All of
the line measurements were made by integrating over the specific line
region (8 \AA\ wide centered on the line) and subtracting off the mean
flux calculated from two adjacent continuum regions (Continuum A and
Continuum B in Table \ref{lines}).  Equivalent widths (EW) were
computed for each line by dividing the integrated line flux by the
mean continuum value \citep[as in previous studies, we define the EW
to be positive for emission lines;][]{hawley96, west04}.  The
low-resolution of SDSS spectra does not affect our ability to measure
accurate line values, with the exception of CaII K.  Due to the
intrinsic CaII K absorption present in M dwarf photospheres, weak
activity appears as an emission peak in the absorption line.  This
weak emission cannot be resolved in SDSS spectra.  However strong CaII
K emission overwhelms the absorption and can be easily recovered from
low-resolution spectra \citep{walkowicz09}.  The lines and continuum
regions (in vacuum wavelengths) can be found in Table \ref{lines}.
Formal uncertainties on the EWs for each line were also computed.

\begin{deluxetable*}{lcccc}
\tablewidth{0pt}
\tablecolumns{5} 
\tabletypesize{\small}
\tablecaption{Emission Lines}
\renewcommand{\arraystretch}{.6}
\tablehead{
\colhead{Line}&
\colhead{Central}&
\colhead{Continuum A}&
\colhead{Continuum B}&
\colhead{Activity Limit}\\
\colhead{}&
\colhead{Wavelength (\AA)}&
\colhead{(\AA)}&
\colhead{(\AA)}&
\colhead{EW (\AA)}}
\startdata
H$\alpha$&6564.66& 6555.0--6560.0 & 6570.0--6575.0 & 0.75\\
H$\beta$&4862.69&  4840.0--4850.0 & 4875.0-4885.0 & 1.00\\
H$\gamma$&4341.69& 4310.0--4330.0  & 4350.0--4370.0 & 1.00 \\
H$\delta$&4102.90&  4075.0--4095.0 & 4110.0--4130.0 &  1.50\\
CaII K&3934.78&   3952.8--3956.0 & 3974.0--3976.0  &  1.50
\enddata
\tablecomments{All of the wavelengths are given in vacuum units}
\label{lines}
\end{deluxetable*}

Similar to the \citet{west04,west08} samples, we defined magnetically
active stars as those that had detectable emission lines in their
spectra.  Previously, this had been done solely for H$\alpha$.  To be
classified as active in a specific emission line, the following four
criteria had to be satisfied: 1) the EW of the line was larger than
some minimum value (see below); 2) the EW value must be 3 times the
uncertainty; 3) the signal to noise ratio (SNR) in the continuum must
be larger than 3; and 4) the height of the spectral line must be
larger than 3 times the noise in the continuum.  Stars were classified
as inactive if they met criterion \#3 and had no detectable emission.
Criterion \#3 preferentially selects brighter stars at any given color
or spectral type.  This selection bias should not have a large effect
our results since we examine the properties of our sample in a
Galactic context and both active and inactive stars with low SNR
should be removed.  However, if active stars are intrinsically
more luminous than their inactive counterparts (at the same color; see
Section 3.2 below), criterion \#3 may serve to include slightly more
active stars in our resulting analyses.

Because the strength of the stellar continuum changes dramatically
from CaII K to H$\alpha$ in M dwarfs, the EW thresholds used in
criterion \#1 are different for each line.  We experimentally derived
the EW thresholds by automatically measuring activity in a subsample
of 300 M dwarfs (that spanned a range of SNR and spectral types) using
10 different EW activity thresholds.  Additionally, each spectrum was
visually inspected and the activity state for all five emission lines
was determined.  By comparing the results of our manual inspection
with the 10 different automatic routines, we determined the EW
thresholds that most accurately reproduced our visual activity
classifications.  The EW activity threshold for each of the five
emission lines can be found in Table \ref{lines}.  We found that the
previous H$\alpha$ activity limit of 1 \AA\ \citep{west04} was
slightly too high and excluded a number of good detections.  We have
therefore reduced the H$\alpha$ EW activity criterion to 0.75 \AA.

Recently, \citet[hereafter K10]{kruse10} investigated the time
variability of SDSS M dwarfs and defined their H$\alpha$ activity
thresholds based on the median EW values that were detected at a
3$\sigma$ confidence level as a function of spectral type.  Although
the \nocite{kruse10}K10 threshold EWs provide a robust way to select a
clean sample, it unduly biases the sample toward the most active stars
at later spectral types.  For example, our DR7 catalog contains a
large number of active M6 dwarfs (that have H$\alpha$ emission lines
with peak values that are more than 5 times larger than the noise)
that have EWs below the 3\AA\ threshold used in the
\nocite{kruse10}K10 study.  In addition, the activity fractions
presented in \nocite{kruse10}K10 (see their Figure 4) include these
active stars (with EW $<$ 3\AA) in the ``inactive'' component of the
activity fraction denominator.  This explains why the activity
fractions reported by \nocite{kruse10}K10 are smaller than previous
determinations \citep{west04,west08} and indicates a serious selection
bias in the \nocite{kruse10}K10 results.

For all of the active stars in the sample we computed the ratio of
luminosity in the emission line as compared to the bolometric
luminosity ($L_{\rm{line}}$/$L_{\rm{bol}}$).  We followed the methods
of \citet{hall96}, \citet{walkowicz04} and \citet{westhawley08} who
derived $\chi$ factors for the Hydrogen Balmer and Ca II chromospheric
lines as a function of M dwarf spectral type.  The $\chi$ factor uses
empirical results for the bolometric luminosities of M dwarfs as a
function of color and/or spectral type \citep{leggett96,leggett01} and
relates the EW of a line to the fraction of the bolometric luminosity
emitted by the line.  The $L_{\rm{line}}$/$L_{\rm{bol}}$ values were
computed by multiplying the EW of each active star by the appropriate
$\chi$ value.  Formal uncertainties were computed for each
$L_{\rm{line}}$/$L_{\rm{bol}}$ value and are included in the final
database.

We also computed the metal sensitive parameter $\zeta$, defined by
\citet{lepine07}, which uses a combination of the TiO5, CaH2, and CaH3
molecular band indices to separate the sample into different
metallicity classes (see Equations 1 \& 2).  This is similar to the
\citet{gizis97} classification system but was re-calibrated using wide
common proper motion pairs that were assumed to be at the same
metallicity.  Stars with solar metallicity ([Fe/H]=0) have $\zeta$
values $\sim$1 and stars with [Fe/H]=-1 have $\zeta$ $\sim$0.4
\citep{woolf09}.  Although there is considerable scatter in the [Fe/H]
versus $\zeta$ relation at high-metallicities, this parameter is very
useful for finding and classifying low-metallicity stars that are
likely members of the Galactic halo.

Our catalog containing all of the measured quantities is publicly
available on the
Vizier\footnote{http://vizier.u-strasbg.fr/cgi-bin/VizieR} site or can
be obtained by contacting the corresponding author.  The individual
spectra are available from the SDSS DR7
website\footnote{http://www.sdss.org/dr7}.  As with previous SDSS
spectroscopic catalogs of low-mass stars, we remind the community that
these data do not represent a complete sample and that the complicated
SDSS spectral targeting (see Section 3.1) introduces a variety of
selection effects.  However, our new
sample covers a large range of values for many of the physical
attributes of the M dwarfs, including parameters that are sensitive to
activity, metallicity, and Galactic motion, making accurate activity,
kinematic, and chemical analyses possible. In addition, because some
of the derived quantities are computed by automatic routines, values
for a small percentage \citep[$\sim$4\%;][]{west04} of individual stars may be incorrect; this
should not affect large statistical results. Users are nevertheless
cautioned to understand the origin of specific data products before
using them indiscriminately.

\section{Results}

\begin{figure}
\plotone{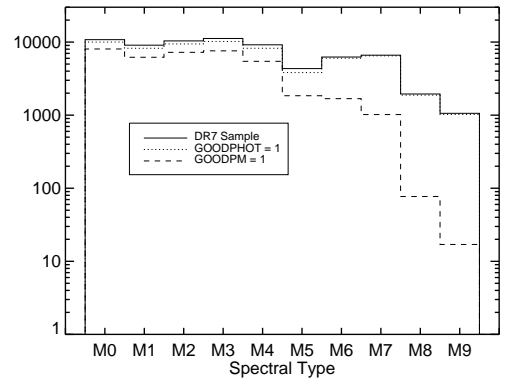}
\caption{The distribution of spectral types in the spectroscopic SDSS
  DR7 M dwarf catalog (solid).  The dotted line represents the stars
  for which good quality photometry exists (GOODPHOT = 1) and the
  dashed line indicates the stars for which good proper motions exist
  (GOODPM =1).  Due to the shallow depths and blue sensitivity of the
  USNO-B catalog, there
  are only a small number of measured proper motions of the latest-type M
  dwarfs.}
\label{hist}
\end{figure}

Figure \ref{hist} shows the distribution of spectral types for the DR7
low-mass star sample.  We included the entire visually confirmed
sample (solid), the stars for which we have good photometry (as
indicated by our GOODPHOT flag; dotted), and the stars for which we
have 3D kinematic information (as set by our GOODPM flag; dashed).  A
majority of the late-type M dwarfs are removed from the kinematic
sample because they are too faint to be detected in the USNO-B
catalog.  Future studies will measure proper motions for the late-type
M dwarfs in the DR7 catalog and create a more representative kinematic sample.

\subsection{Median Colors}

All of the magnitudes included in our DR7 catalog and used to compute
the median colors were extinction corrected using the \citet{sfd}
line-of-sight dust maps \citep[and the dust model of][and $R_V$ =
3.1]{CCM} .  We provide extinction corrections for each band in the
catalog so that observed magnitudes can be easily reproduced.  Because
of the close proximity of many of the M dwarfs in the sample, the
entire \citet{sfd} extinction is likely an overestimate for many
lines-of-sight.  The uncertainty in the extinction correction 
introduces a small uncertainty in the derived distances.  However, given the
small amount of extinction toward objects in the SDSS footprint, the added
uncertainties to the distances are typically less than 5\%.  We are currently
determining the amount of extinction needed to reproduce an unreddened
spectrum for each object in our DR7 sample, and will report on that
analysis in a future paper \citep{jones11}.

We computed median colors for each subtype of the DR7 sample, applying
stringent cuts to include only those stars with the highest quality
photometry.  To be included in the median color calculation, SDSS
stars had to have GOODPHOT = 1, $r$-band extinction $<$0.05
magnitudes, and $r$, $i$ and $z$-band magnitude uncertainties $<$0.05
magnitudes.  Median colors incorporating 2MASS magnitudes were
calculated from the subset of stars with $J$, $H$, and $K_S$ uncertainties
$<$0.05.  The median colors and standard deviations (in parentheses)
are shown in Table \ref{tab:color}.  Due to the systematic offset in
the Hammer automatic spectral types for late-type M dwarfs, our
average colors are bluer than previously reported
\citep[W08,][]{kowalski09}.  We include the $g-r$ colors but warn that
they are likely metallicity dependent \citep[see also Section 3.3
below]{west04, bootem}.

\begin{deluxetable*}{lcccccccc}
\tablewidth{0pt}
\tablecolumns{9} 
\tabletypesize{\scriptsize}
\tablecaption{Median Colors}
\renewcommand{\arraystretch}{.6}
\tablehead{
\colhead{Sp. Type}&
\colhead{$N_{\rm{SDSS}}$\tablenotemark{a}}&
\colhead{$N_{\rm{2MASS}}$\tablenotemark{b}}&
\colhead{$g-r$\tablenotemark{c}}&
\colhead{$r-i$}&
\colhead{$i-z$}&
\colhead{$z-J$}&
\colhead{$J-H$}&
\colhead{$H-K_S$}}
\startdata
M0 &   915 &  259 & 1.31 (0.16) & 0.56  (0.08) &  0.33 (0.06) & 1.20  (0.25) &  0.64  (0.07)  & 0.16  (0.07)\\
M1 &   699 &   230 &  1.39  (0.16) &  0.73  (0.10) &  0.41  (0.08) &  1.32  (0.41) &  0.62  (0.08) &  0.19  (0.08)\\
M2 &   1078 &   689 &   1.40  (0.13) &  0.96  (0.10) &  0.53  (0.08) &  1.26  (0.18) &  0.60  (0.07) &  0.22  (0.08)\\
M3 &   1220 &  831 &  1.41  (0.13) &  1.13  (0.11) &  0.61  (0.08) &  1.30  (0.24) &  0.59  (0.06) &  0.23  (0.06)\\
M4 &   814 &   522 &   1.46  (0.15) &  1.33  (0.13) &  0.71  (0.09) &  1.37  (0.15) &  0.59  (0.06) &  0.25  (0.06)\\
M5 &    328 &   269 &   1.52  (0.13) &  1.62  (0.21) &  0.90  (0.13) &  1.47  (0.12) &  0.59  (0.06) &  0.28  (0.07)\\
M6 &  359 &  401 &  1.56  (0.15) &  1.92  (0.14) &  1.05  (0.07) &  1.61  (0.08) &  0.60  (0.06) &  0.31  (0.08)\\
M7 &   160 &   352 &    1.58  (0.16) &  2.09  (0.19) &  1.14  (0.10) &  1.71  (0.10) &  0.60  (0.08) &  0.34  (0.07)\\
M8 &    6 &    118 &  1.64  (0.12) &  2.56  (0.24) &  1.41  (0.13) &  1.93  (0.13) &  0.63  (0.07) &  0.39  (0.08)\\
M9 &    4 &    77 &  1.87  (0.74) &  2.70  (0.20) &  1.71  (0.18) &  2.15  (0.14) &  0.69  (0.09) &  0.42  (0.07)
\enddata
\tablenotetext{a}{The SDSS numbers reflect the stars that have good SDSS photometry (the GOODPHOT flag set to 1) and have $r$, $i$ and $z$ photometric uncertainties $<$ 0.05 magnitudes.}
\tablenotetext{b}{The 2MASS numbers indicate the number of stars that have $z$, $J$, $H$ and $K_S$-band uncertainties $<$ 0.05 magnitudes.}
\tablenotetext{c}{The $g-r$ colors were calculated using stars with good SDSS photometry (the GOODPHOT flag set to 1) and $g$ and $r$-band photometric uncertainties $<$ 0.05 magnitudes.}
\tablecomments{The median and rms scatter (in parentheses) of the extinction corrected SDSS-2MASS colors for each M dwarf spectral type.  $u-g$ colors were not included due to the large uncertainties in the $u$-band photometry for most of the sample M dwarfs.  The $g-r$ colors are included but are likely metallicity dependent  \citep{west04, bootem}.}
\label{tab:color}
\end{deluxetable*}

\begin{figure}
\plotone{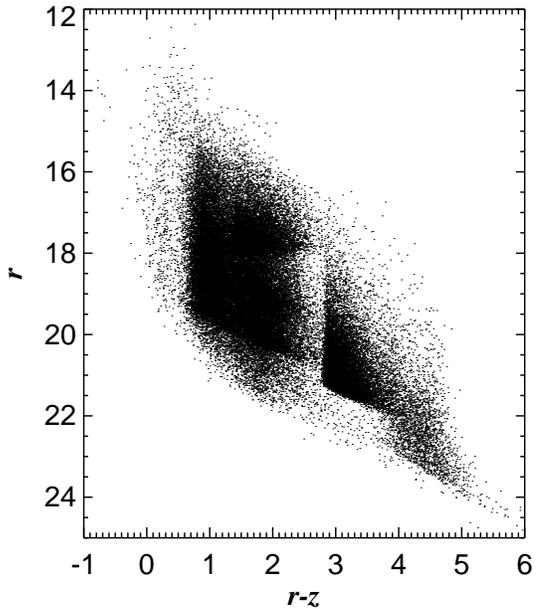}
\caption{The $r$ versus $r-z$ Hess diagram for the M dwarfs in the
  SDSS DR7 sample.  The gaps in color space reflect the non-uniform
  sampling due to the spectral targeting algorithm.  The irregular
  sampling has a minimal effect on the median colors derived in the
  DR7 sample.}
\label{fig:hess}
\end{figure}

Figure \ref{fig:hess} shows the $r$ versus $r-z$ Hess diagram for the
M dwarfs in the SDSS DR7 sample. The gaps in color space reflect the
non-uniform sampling due to the spectral targeting algorithm.  To
quantify the effect of SDSS spectroscopic targeting algorithms on the
median colors, we conducted the following test.  We began with the
straw man assumption that the median $r-i$, $i-z$ and $r-z$
color-spectral type relations reported in Table \ref{tab:color} are
representative of the underlying stellar population.  We then used
these quantities, along with the observed color spread in each
spectral type bin, to construct a synthetic stellar population with a
uniform distribution in both magnitude and spectral type.  We then
applied photometric cuts to the smooth underlying distribution,
replicating the SDSS spectroscopic selection and producing a synthetic
``observed'' sample that is highly structured in color-magnitude
space. The median colors and spreads were then calculated as a
function of spectral type from this synthetic sample and compared to
the measured colors in Table \ref{tab:color}. In general, the colors
of the synthetic stars agree to within 0.02 mags of the observed
sample (an agreement that would not be expected if the underlying
population were different from the observed SDSS stars ) and we
conclude that the colors presented in Table \ref{tab:color} are not
severely affected by the spectroscopic selection.

\subsection{Activity}

\begin{figure}
\plotone{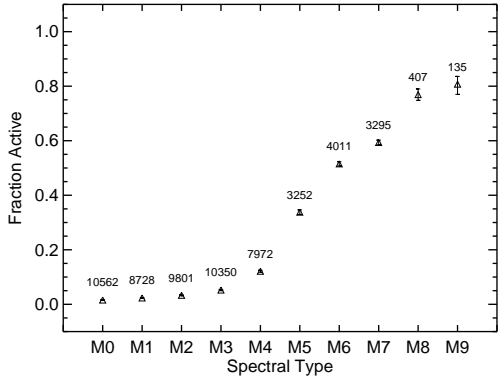}
\caption{The H$\alpha$ magnetic activity fraction as a function of
  spectral type.  Error bars were calculated from the binomial
  distribution and the number above each data point indicates the
  numbers of stars in that bin. See text for more details.}
\label{hafrac}
\end{figure}

Figure \ref{hafrac} shows the H$\alpha$ activity fraction as a
function of M dwarf spectral type for stars with data quality
sufficient to measure H$\alpha$ emission (and GOODPHOT =1 and WDM =0).
Our results are in good agreement with previous studies
\citep[W08]{west04}, but with much lower uncertainties due to the
$\sim$59,000 M dwarfs used to generate Figure \ref{hafrac} (7952 are
H$\alpha$ active and 51,034 are H$\alpha$ inactive).  There are a few
small differences between Figure \ref{hafrac} and the H$\alpha$
activity fractions reported by \nocite{west08}W08 that are likely due to the
changes in some of the spectral types (see above) and the different
Galactic sightlines included in the DR7 sample.  Because of the lower
SNR in the blue portion of M dwarfs, fewer stars were available to
measure activity for the higher order Balmer and CaII K emission
lines.  Table \ref{activity} gives the number of active and inactive
stars for each emission line indicator that had the necessary quality
for our activity analysis.

As discussed in \nocite{west08}W08, the activity fraction is highly
dependent on the location of the samples in the
Galaxy.  Because M dwarfs have finite activity lifetimes and are
dynamically heated in the Galactic disk as they age, their activity
state is correlated with position in the Galaxy.
One of the reasons that the M dwarfs in our sample have much smaller
activity fractions than those studied nearby
\citep[e.g.][]{hawley96,gizis00}, is that the SDSS volume is much
larger than those used in previous catalogs and is concentrated on the north
galactic cap, yielding a much older stellar population.  This is
particularly clear for the early-type M dwarfs, which have median
distances greater than 500 pc, and therefore ages that are
considerably older than their short activity lifetimes ($\sim$ 1-2
Gyr).  We reiterate the warning from \nocite{west08}W08 that activity
fractions in M dwarfs must be discussed in the
proper Galactic context.

\begin{deluxetable}{lccc}
\tablewidth{0pt}
\tablecolumns{4} 
\tabletypesize{\small}
\tablecaption{Activity Indicators}
\renewcommand{\arraystretch}{.6}
\tablehead{
\colhead{Line}&
\colhead{$N_{\rm{active}}$}&
\colhead{$N_{\rm{inactive}}$}&
\colhead{Active Fraction}}
\startdata
H$\alpha$&7952&51034&0.13\\
H$\beta$&2236&35144&0.06\\
H$\gamma$&1175&18816&0.06\\
H$\delta$&528&14664&0.03\\
CaII K&620&9053&0.06
\enddata
\label{activity}
\end{deluxetable}

\begin{figure*}
\plotone{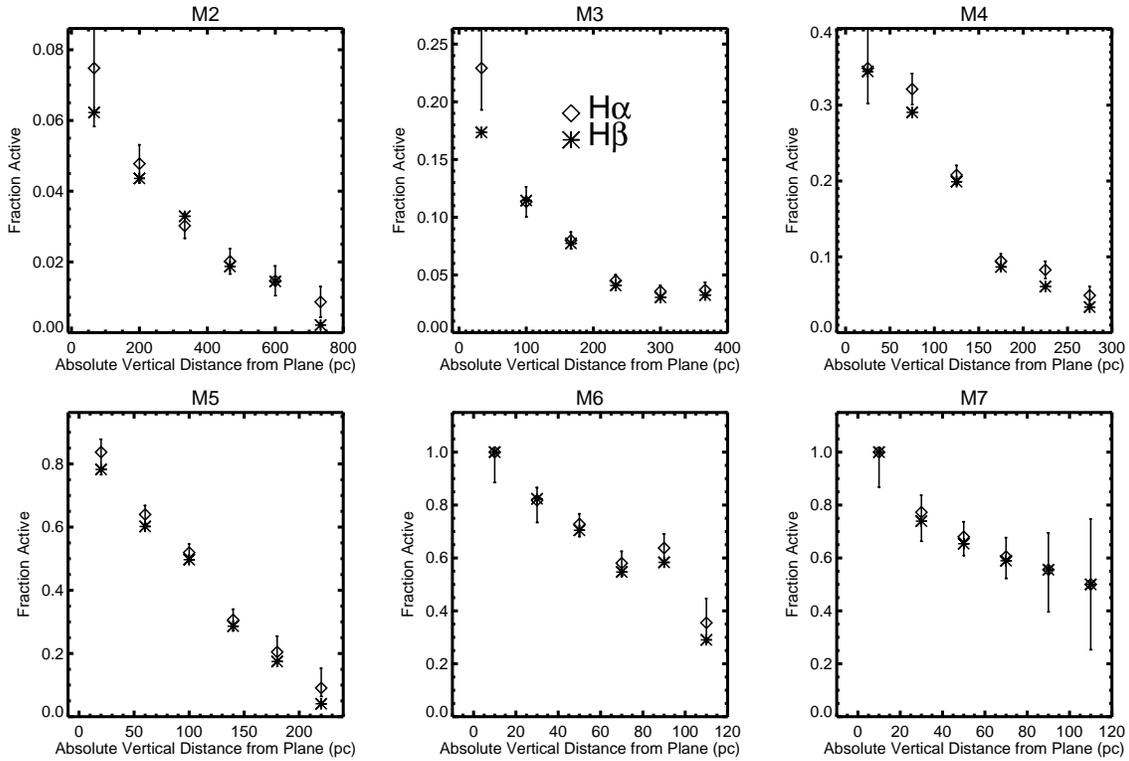}
\caption{The H$\alpha$ and H$\beta$ activity fractions as a function
  of absolute vertical distance from the Galactic plane for M2--M7
  dwarfs.  For this comparison, only the stars with data good enough to measure
  H$\beta$ were included.  Error bars indicate the binomial errors in
  the H$\alpha$ fractions -- the H$\beta$ uncertainties are similar.
  The H$\beta$ activity fractions are in excellent agreement with the
  H$\alpha$ data, indicating that H$\beta$ activity has the same duration in M
  dwarfs.}
\label{betafrac}
\end{figure*}

The various emission lines measured in our spectra are formed at
slightly different locations in the chromosphere, suggesting that the
strength of one emission line may not necessarily predict the strength
of another.  We therefore examined how the various activity induced
emission lines trace each other as a function of spectral type and
absolute distance from the Galactic plane (a proxy for age).  Figure
\ref{betafrac} shows the H$\alpha$ (diamonds) and H$\beta$ (asterisks)
activity fractions for M2-M7 dwarfs as a function of Galactic height.
Only the H$\alpha$ activity fractions for stars that could have
detected H$\beta$ emission are included.  This limits the sample to
the brighter stars at each spectral type that meet the SNR criterion
and does not include all of the M dwarfs that are sensitive to
H$\alpha$ emission.  Spectral types earlier than M2 and later than M7
were not included due to an insufficient sample size of active stars.  As seen in
previous studies, the magnetic activity fractions decrease as a
function of Galactic height, confirming an age-activity relationship;
stars closer to the Plane are statistically younger and more likely to
be active.  Figure \ref{betafrac} also demonstrates that H$\alpha$ and
H$\beta$ trace each other extremely well for the entire DR7 sample,
implying that the H$\beta$ activity lifetimes are essentially the same
as those for H$\alpha$.

\begin{figure*}
\plotone{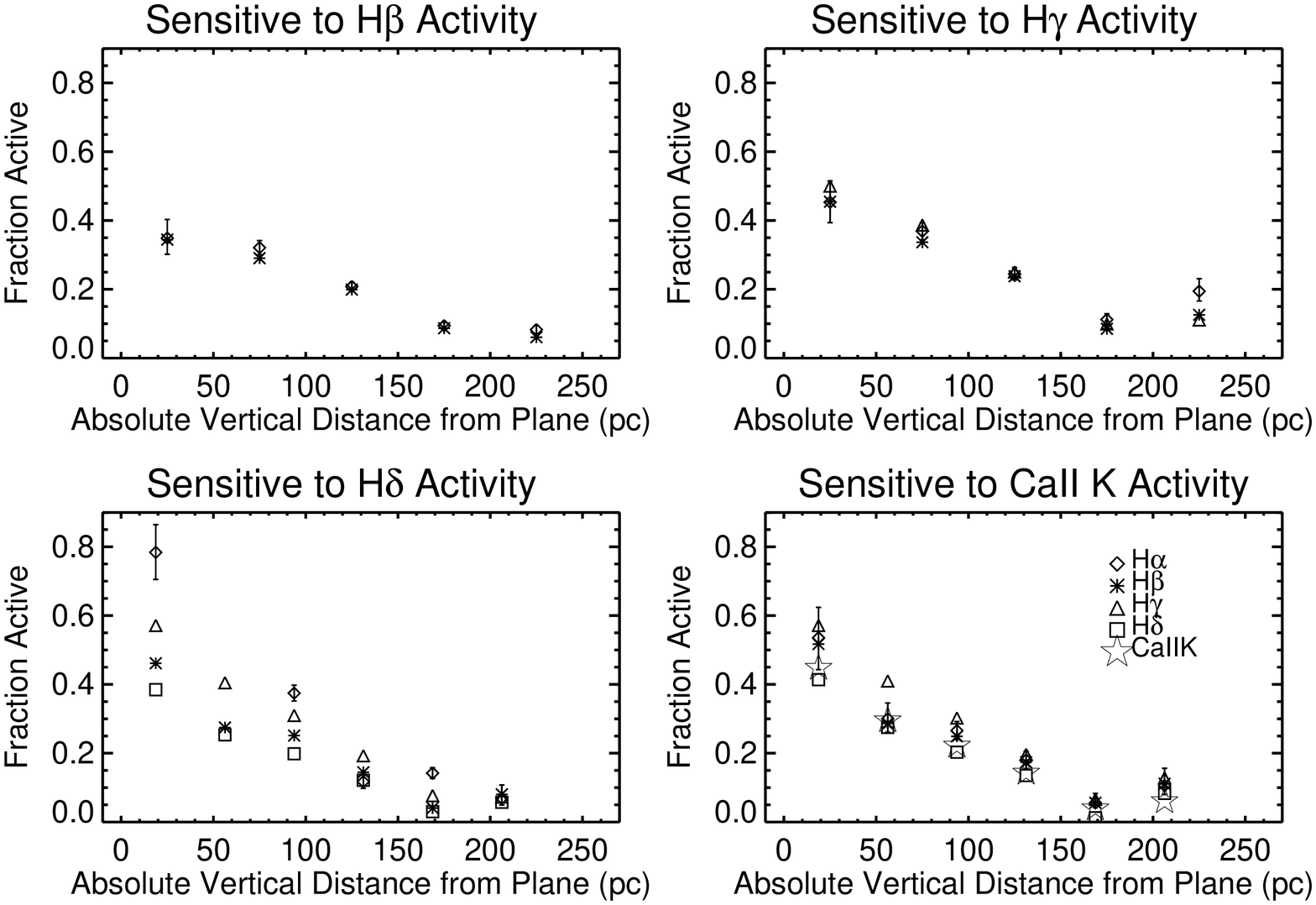}
\caption{The H$\alpha$, H$\beta$, H$\gamma$, H$\delta$ and CaII K
  activity fractions for M4 dwarfs as a function of absolute vertical
  distance from the Galactic plane.  Each panel includes only the
  stars where the data are good enough to be sensitive to that
  emission line (as well as all of the redder emission lines).
  Uncertainties are included for H$\alpha$ but are similar for all of
  the emission lines.  The activity fractions have the same magnitude
  and decreasing trend with Galactic height for all lines.  The
  activity fractions for stars sensitive to H$\delta$ and CaII K
  activity are slightly higher than for stars sensitive to H$\beta$
  and H$\gamma$.}
\label{m4frac}
\end{figure*}

The other activity tracers also appear to correlate with H$\alpha$
emission.  Figure \ref{m4frac} shows the activity fractions for all
five emission lines in M4 dwarfs as a function of vertical distance
from the Galactic plane.  Each panel plots the activity fractions for
stars where the data are good enough to measure (i.e. are sensitive
to) H$\beta$ (top left), H$\gamma$ (top right), H$\delta$ (bottom
left) and CaII K (bottom right) activity as well as the emission lines
redder than that specific tracer.  The activity fractions for all of
the lines are in excellent agreement with each other and indicate that
all five emission lines are produced for the same amount of time
during the lifetime of an M dwarf.  This new result is complementary
to previous studies that found strong correlations among different
emission line strengths in M dwarfs \citep{rm06,walkowicz09}.
Although the relative emission line strengths for individual stars may
fluctuate over time \citep[see][]{cincunegui07}, the mean values of
the emission lines are well correlated in the large DR7 sample.  The
unprecedented size and Galactic distribution of our spectroscopic
sample confirms that not only the strength, but the duration of
activity appears to be similar for all of the Balmer and CaII activity
indicators.

Figure \ref{m4frac} also demonstrates a possible (and important)
selection effect present in magnitude-limited or activity-selected
samples.  A close examination of Figure \ref{m4frac} (and Figure
\ref{betafrac}) reveals that the activity fractions are larger for the
stars that are sensitive to the bluer emission lines.  This trend was
not expected since the stars used to compute the activity fraction were
drawn from the same volumes in each of the four panels.  The only
difference was that the stars selected for the bluer emission line
analyses were required to have higher SNR.  With this in mind, we can
explain the different activity fractions as follows.

The actual spread in absolute magnitude (or luminosity) at a given
spectral type can be quite significant.\footnote{Recent studies have
  confirmed that the spread in absolute magnitude as a function of
  color is much smaller \citep{boo10, boostat, faherty11}.  This is
  the primary reason that our distances are estimated from colors and
  not spectral types.}  Some of this spread may be due to differences
in the physical properties of the stars.  \citet{hawley96} showed that
magnetically active stars are brighter in $M_V$ than their inactive
counterparts.  Recently, \citet[hereafter Paper II]{boostat}
found that both active and higher metallicity M dwarfs (many stars are
both) appear to be brighter in $M_r$ at a given color or spectral
type. Because the bluer emission lines that we are examining in Figure
\ref{m4frac} require higher SNR spectra to accurately measure
activity, the M4 dwarfs at a given distance that meet the SNR
criterion are preferentially more luminous than other stars in the
same distance bin.  If activity is correlated with luminosity, our SNR
selection will bias our sample toward containing a higher fraction of
active stars.

In addition to activity fractions, we also investigated the median
luminosity in each emission line as a function of spectral
type.  Previous studies have investigated how the H$\alpha$ luminosity
changes as a function of spectral type
\citep{hawley96,gizis00,burgasser02,west04}.  These studies found that
the fraction of luminosity emitted in H$\alpha$ is constant over the
range M0-M5 and decreases by almost an order of magnitude for
late-type M dwarfs.  However, most of the previous studies
concentrated on H$\alpha$ and did not include similar analyses for the
other Balmer or CaII lines, and those that did were based on a small
sample of stars \citep{hawley91,westhawley08}.

\begin{figure*}
\epsscale{0.60}
\plotone{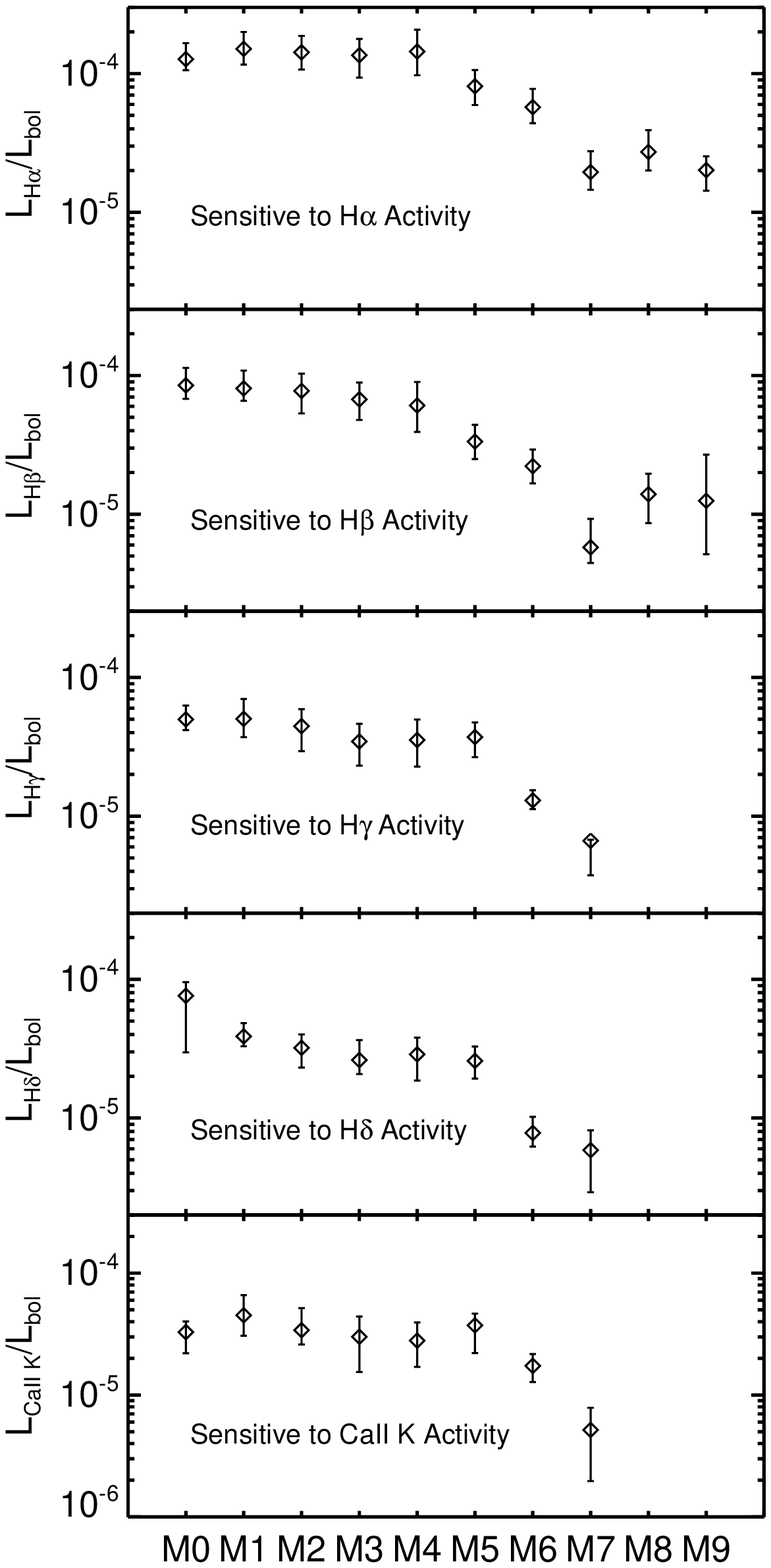}
\caption{The ratio of line luminosity to bolometric luminosity as a
  function of spectral type for H$\alpha$, H$\beta$, H$\gamma$,
  H$\delta$ and Ca II K.  The higher order Balmer and CaII K emission lines
  have progressively lower luminosity than the H$\alpha$ lines.  All
  of the emission lines show a drop in the luminosity ratio for
  late-type M dwarfs.}
\label{lbol}
\end{figure*}

Each panel of Figure \ref{lbol} shows the median luminosity of a
specific activity tracer (as compared to the bolometric luminosity of
the stars) for active stars sensitive to that particular emission
line.  All of the activity tracers exhibit similar behavior, with the
relative luminosity staying constant for the early-type M dwarfs and
then falling to lower values at or around a spectral type of M5.  The
median values and the upper and lower quartiles that are plotted in
Figure \ref{lbol} are also given in Table \ref{tab:lbol}. The bluer
emission lines produce smaller fractional luminosities as expected,
and are consistent with previous Balmer decrement studies
\citep{hawley91,bootem,westhawley08}. \nocite{west08}W08 found evidence
for a slight decrease in L$_{\rm{H\alpha}}$/L$_{\rm{bol}}$ as a
function of Galactic height, but the effect is small and not included
in our analysis.

\begin{deluxetable*}{cccccc}
\tablewidth{0pt}
\tablecolumns{6} 
\tabletypesize{\small}
\tablecaption{Luminosity in Activity Tracer Emission Lines}
\renewcommand{\arraystretch}{1}
\tablehead{
\colhead{Spectral}&
\colhead{$L_{\rm{H\alpha}}$/L$_{\rm{bol}}$}&
\colhead{$L_{\rm{H\beta}}$/L$_{\rm{bol}}$}&
\colhead{$L_{\rm{H\gamma}}$/L$_{\rm{bol}}$}&
\colhead{$L_{\rm{H\delta}}$/L$_{\rm{bol}}$}&
\colhead{$L_{\rm{CaII\ K}}$/L$_{\rm{bol}}$}\\
\colhead{Type}&
\colhead{($\times10^{-4}$)}&
\colhead{($\times10^{-4}$)}&
\colhead{($\times10^{-4}$)}&
\colhead{($\times10^{-4}$)}&
\colhead{($\times10^{-4}$)}}
\startdata
M0 &     1.27$_{-0.21}^{+0.39}$ &   0.85$_{-0.17}^{+0.29}$ &    0.50$_{-0.08}^{+0.13}$ &    0.76$_{-0.46}^{+0.19}$ &    0.33$_{-0.11}^{+0.07}$\\
M1  & 1.51$_{-0.35}^{+0.49}$ &    0.81$_{-0.15}^{+0.28}$ &    0.50$_{-0.13}^{+0.20}$ &    0.39$_{-0.06}^{+0.09}$ &    0.45$_{-0.14}^{+0.21}$\\
M2 &  1.42$_{-0.35}^{+0.45}$ &    0.77$_{-0.24}^{+0.26}$ &    0.44$_{-0.15}^{+0.15}$ &    0.32$_{-0.09}^{+0.08}$ &    0.34$_{-0.08}^{+0.18}$\\
M3  & 1.36$_{-0.42}^{+0.42}$ &    0.67$_{-0.19}^{+0.22}$ &    0.35$_{-0.11}^{+0.12}$ &    0.26$_{-0.05}^{+0.10}$ &    0.30$_{-0.15}^{+0.14}$\\
M4  & 1.44$_{-0.47}^{+0.63}$ &    0.61$_{-0.22}^{+0.29}$ &    0.35$_{-0.13}^{+0.14}$ &    0.29$_{-0.10}^{+0.09}$ &    0.28$_{-0.11}^{+0.11}$ \\
M5  &  0.81$_{-0.22}^{+0.25}$ &    0.33$_{-0.08}^{+0.11}$ &    0.37$_{-0.11}^{+0.10}$ &    0.26$_{-0.07}^{+0.07}$ &    0.37$_{-0.15}^{+0.09}$\\
M6  &  0.57$_{-0.13}^{+0.20}$ &    0.22$_{-0.06}^{+0.07}$ &    0.13$_{-0.02}^{+0.02}$ &    0.08$_{-0.02}^{+0.02}$ &    0.17$_{-0.05}^{+0.04}$\\
M7   & 0.19$_{-0.05}^{+0.08}$ &    0.06$_{-0.01}^{+0.03}$ &    0.07$_{-0.03}^{+0.00}$ &    0.06$_{-0.03}^{+0.02}$ &    0.05$_{-0.03}^{+0.03}$\\
M8   & 0.27$_{-0.07}^{+0.12}$ &    0.14$_{-0.05}^{+0.06}$ &    \nodata & \nodata & \nodata\\
M9   &  0.20$_{-0.06}^{+0.05}$ &    0.12$_{-0.07}^{+0.14}$ &    \nodata & \nodata & \nodata\\
\enddata
\label{tab:lbol}
\tablecomments{Bolometrically scaled luminosities were derived using stars that show activity in each emission line.  Reported uncertainties are the 25th and 75th quartile values.}
\end{deluxetable*}

\subsection{Metallicity Sensitive Features}

\begin{figure}
\plotone{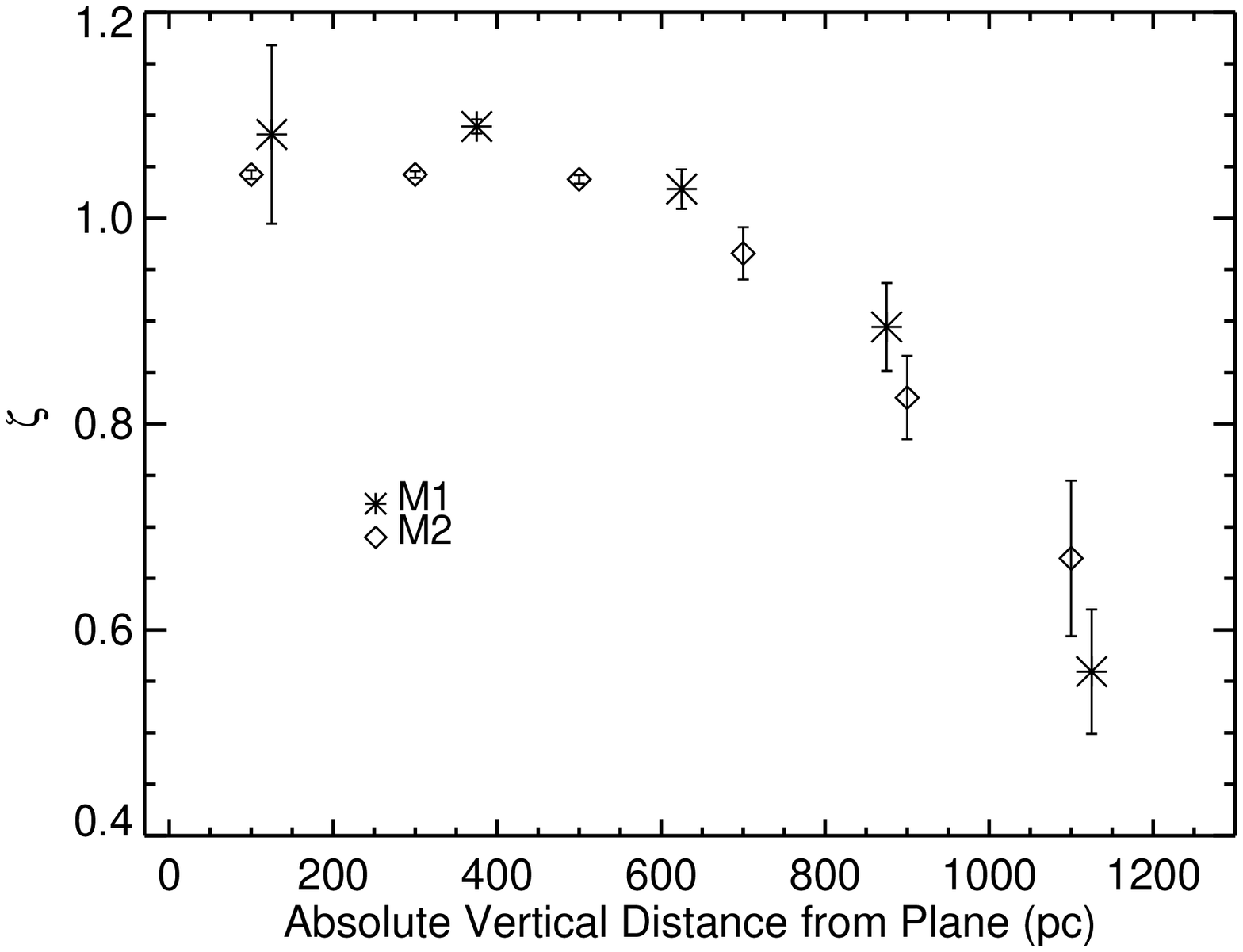}
\caption{The metallicity dependent parameter $\zeta$ as a function of
  absolute height above the Galactic plane for M1 and M2 dwarfs.
  Metallicity decreases at large Galactic heights uniformly for both
  spectral types, indicating a significant evolution of M dwarf
  metallicity in the Galactic disk.  The error bars reflect the
  uncertainty in the median values in each bin.  This trend, along
  with the ubiquity and longevity of M dwarfs, paves the way for
  future studies of the chemical evolution of the nearby Milky Way.}
\label{zeta}
\end{figure}

Metallicity continues to be one of the most elusive physical
quantities to measure in M dwarfs.  Recent work using infrared
spectroscopy and improved photometry for M dwarfs in wide binaries may
soon lead to accurate metallicity determinations for large samples of
M dwarfs \citep{woolf09,johnson09,babs10}.  However, the
\citet{lepine07} $\zeta$ parameter is currently the best indicator of
metallicity in optical M dwarf spectra.  Using the DR7 M dwarf sample,
we examined how $\zeta$ changes as a function of Galactic height.
Figure \ref{zeta} shows median $\zeta$ values for M1 and M2 dwarfs as
a function of absolute height above the Galactic plane.  These
specific spectral types were chosen because they span a large range of
distance and contain a significant number of stars with $\zeta$
uncertainty $<$ 0.1. Figure \ref{zeta} shows a clear decrease in
$\zeta$ for both spectral types with increasing Galactic height.  The
discrepancies in the nearby bins reflect the scatter in the
high-metallicity end of the $\zeta$ vs. metallicity relation
\citep{woolf09}.  The results shown in Figure \ref{zeta} suggest a
significant change in the median metallicity of M dwarfs (and by
extension the Galaxy) as a function of age, with the oldest stars
(that are farther from the Plane) having the lowest metallicity.
Similar trends have been seen in both the radial and vertical
directions from previous spectroscopic samples of higher mass stars
\citep{nordstrom04, ivezic08}.  However, the higher density and longer
main sequence lifetimes of M dwarfs may provide a better probe of the
metal content (and evolution) of the local Galactic disk.  The $\zeta$
values of $\sim$0.6 in the most distant bins of Figure \ref{zeta} have
metallicities of [Fe/H]$\sim$-0.7 according to the wide binary
analysis of \citet{woolf09}.\footnote{The \citet{woolf09} [Fe/H] --
  $\zeta$ calibration was limited to $\sim$M0-M3 dwarfs and has
  considerable scatter at the high-metallicity end of the relation.}
We note that the photometric distances derived in our sample are
likely affected by metallicity and that the farthest bins in Figure
\ref{zeta} (that have lower metallicities) are closer than shown due
to fainter $M_r$ magnitudes (Paper II)\nocite{boostat}.  Regardless,
this strong trend in metallicity demonstrates the utility of large
spectroscopic samples of M dwarfs (such as the one presented in this
paper) for understanding the chemical evolution of the Milky Way.
However, additional calibration is required to convert quantities such
as $\zeta$ into precise metallicities.

\begin{figure}
\plotone{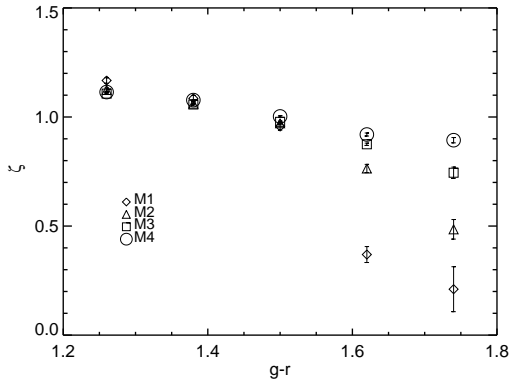}
\caption{The metallicity sensitive parameter $\zeta$ as a function of
  $g-r$ color for M1-M4 stars.  While $\zeta$ decreases at redder
  $g-r$ colors, the decrease is not uniform across spectral type.}
\label{zetagr}
\end{figure}

\begin{figure*}
\plotone{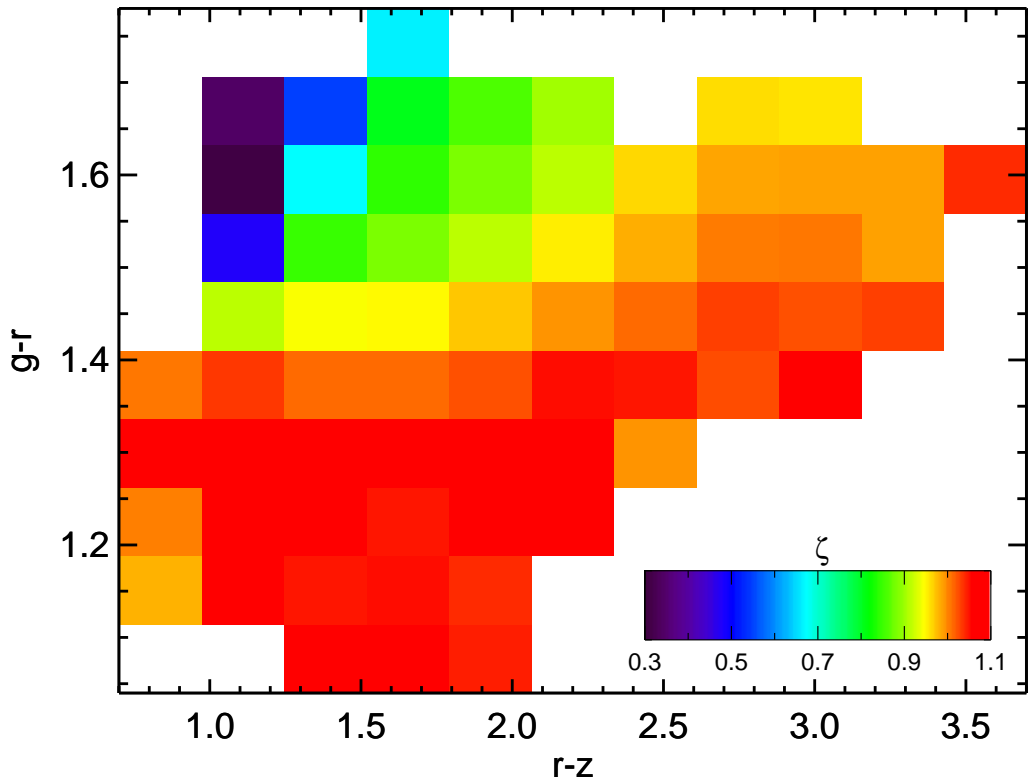}
\caption{The $g-r$ versus $r-z$ colors of M dwarfs in the DR7
  spectroscopic sample.  The bins have been color coded according to
  their $\zeta$ values.  Metallicity decreases along roughly diagonal
  lines in the $g-r$, $r-z$ color space and perpendicular to the
  stellar locus \citep{covey07}.  This relation should be useful for
  classifying M dwarfs and identifying low-metallicity subdwarfs in
  upcoming large photometric surveys.}
\label{zetacol}
\end{figure*}

Previous SDSS studies have shown that the $g-r$ color of M dwarfs
appears to correlate with metallicity \citep{west04, lepine08}.
Future large photometric surveys such as LSST \citep{lsst} will rely
heavily on the ability to classify objects based on their broadband
colors alone.  A large spectroscopic sample of M dwarfs, like the one
presented here, can help correlate photometric properties of M dwarfs
to spectroscopically derived physical parameters. Figure \ref{zetagr}
shows the $g-r$ colors of M dwarfs as a function of $\zeta$ for M1-M4
dwarfs.  Although there is a trend of decreasing $\zeta$ with redder
$g-r$ color, the change in $g-r$ is different across spectral types.
Therefore, $g-r$ color alone is not a good tracer of M dwarf
metallicity across the range of M dwarf spectral types.  However, when
we examined the $\zeta$ values as a function of both $g-r$ and $r-z$
(a proxy for spectral type; see Table \ref{tab:color}), we found a
strong trend of decreasing $\zeta$ diagonal to the $g-r$, $r-z$ axis
and perpendicular to the stellar locus \citep{covey07}.  Figure
\ref{zetacol} highlights a region of color space that can be used to
probe the metallicity content of M dwarfs and obtain low-metallicity
members of the Galactic halo from solely broadband photometry.  Using
a Levenberg-Marquardt minimization technique, we derived a
two-dimensional fit that relates $\zeta$ to the $g-r$ and $r-z$
colors:

\begin{eqnarray}
\zeta= & 1.04-0.98(g-r)^2-0.07(r-z)^2+1.07(g-r)\nonumber \\
 & -0.53(r-z)+0.63(g-r)(r-z).
\end{eqnarray}

\noindent Equation 3 is valid over the ranges shown in Figure \ref{zetacol} and has typical uncertainties of 10-20\%.

\subsection{Nearby Stars}

Prompted by the discovery of new low-mass stars in the solar vicinity
from other SDSS studies \citep[e.g.][]{schmidt_blue}, we inspected our
catalog for M dwarfs that are possibly within 25 pc of the Sun.  We
found 21 nearby M dwarf candidates (8 of which were previously
unidentified) based on the distances derived using the $M_r$,  $r-z$
photometric parallax relation of \citet{boo10}.  Three of the
stars were in the \nocite{west08}W08 SDSS DR5 sample but were not
identified there as potential solar neighbors.  Table \ref{table:dist} gives
the positions, spectral types and distance estimates for all 21 of the
nearby candidates.  The closest candidate is SDSS1410+1846, which has
an estimated distance of 14.8 pc.  Since roughly one out of every
1000 M dwarfs in SDSS has a spectrum, there is a strong possibility
that many more nearby stars await discovery and will be cataloged in
future studies.

\begin{deluxetable*}{lccrrrrrrr}
\tablewidth{0pt}
\tablecolumns{10} 
\tabletypesize{\small}
\tablecaption{M Dwarf Candidates Within 25 pc}
\renewcommand{\arraystretch}{.6}
\tablehead{
\colhead{}&
\colhead{Distance\tablenotemark{a}}&
\colhead{Spectral}&
\multicolumn{3}{c}{R.A. (J2000)}&
\multicolumn{3}{c}{Dec. (J2000)}&
\colhead{}\\
\colhead{Name}&
\colhead{(pc)}&
\colhead{Type}&
\colhead{$^h$}&
\colhead{$^m$}&
\colhead{$^s$}&
\colhead{$^{\circ}$}&
\colhead{$^{\prime}$}&
\colhead{$^{\prime\prime}$}&
\colhead{ref.}}
\startdata
SDSS0013$-$0025 & 23.6 &  M7& 00 & 13 & 09.33 & $-$00 & 25 & 52.0 & 1, 4\\
SDSS0124$-$0027 & 22.8 & M7 & 01 & 24 & 31.25 & $-$00 & 27 & 56.3 & 7\\
SDSS0312+0021 & 24.9 & M7 & 03 & 12 & 25.13 & 00 & 21 & 58.4 & new\\
SDSS0351$-$0052 & 13.5 & M7 & 03 & 51 & 00.03 & $-$00 & 52 & 45.9 & 6\\
SDSS0830+0947 & 20.8 & M9 & 08 & 30 & 32.37 & 09 & 47 & 12.7 & 10\\
SDSS0902+0033 & 20.2 & M8 & 09 & 02 & 06.91 & 00 & 33 & 19.4 & 1\\
SDSS0911+2248 & 21.9 & M7 & 09 & 11 & 30.54 & 22 & 48 & 10.8 & new\\
SDSS1003$-$0105 & 21.2 & M8 & 10 & 03 & 19.15 & $-$01 & 05 & 08.0 & 1, 8\\
SDSS1016+2751 & 18.3 & M8 & 10 & 16 & 34.61 & 27 & 51 & 46.5 & 9, 10\\
SDSS1119+0820 & 24.7 & M8 & 11 & 19 & 46.53 & 08 & 20 & 35.1 & 1, 2\\
SDSS1134+2046 & 16.5 &  M6 & 11 & 34 & 15.72 & 20 & 46 & 54.2 & 5\\
SDSS1252+0252 & 23.5 & M8 & 12 & 52 & 22.63 & 02 & 52 & 05.7 & 1, 2\\
SDSS1253+4034 & 21.1 & M7 & 12 & 53 & 12.49 & 40 & 34 & 00.6 & 1, 9, 10\\
SDSS1326+5640 & 18.0 & M7 & 13 & 26 & 16.31 & 56 & 40 & 44.7 & 1 \\
SDSS1336+4751 & 23.9 & M8 & 13 & 36 & 50.50 & 47 & 51 & 32.2 & 1\\
SDSS1410+1846 & 14.8 & M6 & 14 & 10 & 10.41 & 18 & 46 & 12.0 & new\\
SDSS1422+2116 & 21.7 & M8 & 14 & 22 & 24.27 & 21 & 16 & 07.6 & new\\
SDSS1440+1339 & 24.1 & M7 & 14 & 40 & 22.87 & 13 & 39 & 20.8 & 10\\
SDSS1500$-$0039 & 21.5 & M7 & 15 & 00 & 26.35 & $-$00 & 39 & 27.9 & new\\
SDSS1501+2250 & 19.9 & M9 & 15 & 01 & 08.17 & 22 & 50 & 01.8 & 11, 12\\
SDSS1627+3538 & 24.7 & M7 & 16 & 27 & 18.20 & 35 & 38 & 35.7 & 1, 2, 3\\
\enddata
\label{table:dist}
\tablecomments{(1) \nocite{west08}W08; (2) \citet{lspm}; (3) \citet{westbasri09}; (4) \citet{branham03}; (5) \citet{nltt}; (6) \citet{gliese91}; (7) \citet{salim03}; (8) \citet{deacon05}; (9) \citet{cruz03} ; (10) \citet{schmidt07}; (11) \citet{berger08}; (12) \citet{dahn02}}
\tablenotetext{a}{Photometric distances estimated using the $M_r$, $r-z$ relation from \citet{boo10}.}
\end{deluxetable*}

\section{Conclusions}

We have presented the SDSS DR7 M dwarf spectroscopic catalog, which
consists of more than 70,000 visually confirmed M dwarfs and
represents the largest spectroscopic sample of M dwarfs ever
assembled.  Our value-added catalog includes proper motions, RVs,
photometric matches to 2MASS, spectral classification, distances,
calculated space velocities, activity-induced emission line
measurements and molecular bandhead strengths.  The DR7 catalog is
available for download at the Vizier site or by contacting the
corresponding author.  Our analysis of the visual spectral
classification as compared to the automatic Hammer results reveals a
slight $\sim0.4$ subtype systematic offset in the automatic Hammer spectral types for
late-type M dwarfs but confirms that the automatic types are good to
within the stated precision of 1 spectral subtype.  We present updated median colors for all M
dwarf spectral types.

We have also analyzed some of the bulk
properties of the DR7 sample.  The main results of our analysis are as
follows:

\begin{enumerate}

\item The magnetic activity fractions of low-mass stars as traced by H$\beta$,
  H$\gamma$, H$\delta$, and CaII K, decrease as a function of Galactic
  height and agree with those previously traced using
  H$\alpha$. This confirms that the presence and duration of magnetic
  activity as traced by the higher order Balmer emission lines and CaII K is
  similar to that of H$\alpha$, albeit at lower line luminosities and
  only measured in stars with sufficient SNR.

\item The metallicity sensitive parameter $\zeta$ decreases as a
  function of Galactic height, confirming a decline in the metal
  content of distant M dwarfs (and the Galactic disk) as a function of
  age.

\item As previously shown, the $g-r$ color of M dwarfs correlates with
  metallicity (here parameterized by $\zeta$).  However, the $g-r$
  versus $\zeta$ relation \emph{is} spectral type dependent.  Using
  both the $g-r$ and $r-z$ colors, we demonstrate that low-metallicity
  subdwarfs can be identified using photometry alone.  This relation
  will be useful for source identification in upcoming large
  photometric surveys such as Pan-STARRS and LSST.

\item The DR7 M dwarf sample contains several previously unidentified
  M dwarfs that are likely within 25 pc of the Sun, including one that
  is possibly closer than 15 pc.  The sparse spectroscopic coverage of
  low-mass stars in SDSS suggests that there are numerous M dwarf
  solar neighbors that will be identified in future studies.  The
  advent of large, multi-epoch, deep surveys will be particularly
  useful for completing the nearby M dwarf census.

\end{enumerate}

Many additional studies will make use of our M dwarf catalog.  Those
already underway include a statistical parallax analysis of absolute
magnitude variations in the M dwarf population (Paper
II)\nocite{boostat}, a detailed examination of M dwarf kinematics and the
motions of the Galactic thin and thick disks \citep[Paper
III]{pineda10}, and an investigation of the content and distribution
of dust in the local Galaxy \citep{jones11}.  In addition to the
current and unforeseen science that will be accomplished with our
sample, we anticipate that our M dwarf catalog will be used to select
and classify M dwarfs in several upcoming large surveys.  Over 1
billion M dwarfs will be observed and cataloged in the new wave of
large photometric surveys coming online in the next decade.  We hope
that our large M dwarf sample presented in this paper will provide a
useful tool for correlating the spectroscopic attributes of low-mass
stars with their photometric properties both in single exposures and
in the time domain.

\acknowledgments

The authors would like to thank Sebastian L{\'e}pine, Jackie Faherty,
Adam Burgasser, Evgenya Shkolnik and Edo Berger for useful discussions
leading to the completion of this catalog.  The authors also thank the
anonymous referee for his/her insightful comments, which greatly
improved the quality of the original manuscript.  S.L.H. and
J.J.B. acknowledge the support of NSF AST grant 06-07644.
K.R.C. acknowledges support for this work from the Hubble Fellowship
Program, provided by NASA through Hubble Fellowship grant
HST-HF-51253.01-A awarded by the STScI, which is operated by the AURA,
Inc., for NASA, under contract NAS 5-26555.

Funding for the Sloan Digital Sky Survey (SDSS) and SDSS-II has been
provided by the Alfred P. Sloan Foundation, the Participating
Institutions, the National Science Foundation, the U.S. Department of
Energy, the National Aeronautics and Space Administration, the
Japanese Monbukagakusho, and the Max Planck Society, and the Higher
Education Funding Council for England. The SDSS Web site is
http://www.sdss.org/.

The SDSS is managed by the Astrophysical Research Consortium (ARC) for
the Participating Institutions. The Participating Institutions are the
American Museum of Natural History, Astrophysical Institute Potsdam,
University of Basel, University of Cambridge, Case Western Reserve
University, The University of Chicago, Drexel University, Fermilab,
the Institute for Advanced Study, the Japan Participation Group, The
Johns Hopkins University, the Joint Institute for Nuclear
Astrophysics, the Kavli Institute for Particle Astrophysics and
Cosmology, the Korean Scientist Group, the Chinese Academy of Sciences
(LAMOST), Los Alamos National Laboratory, the Max-Planck-Institute for
Astronomy (MPIA), the Max-Planck-Institute for Astrophysics (MPA), New
Mexico State University, Ohio State University, University of
Pittsburgh, University of Portsmouth, Princeton University, the United
States Naval Observatory, and the University of Washington.

This publication makes use of data products from the Two Micron All
Sky Survey, which is a joint project of the University of
Massachusetts and the Infrared Processing and Analysis
Center/California Institute of Technology, funded by the National
Aeronautics and Space Administration and the National Science
Foundation.

\bibliographystyle{apj}



\begin{thebibliography}{76}
\expandafter\ifx\csname natexlab\endcsname\relax\def\natexlab#1{#1}\fi

\bibitem[{{Abazajian} {et~al.}(2009)}]{dr7}
{Abazajian}, K.~N., {et~al.} 2009, \apjs, 182, 543

\bibitem[{{Berger} {et~al.}(2008){Berger}, {Gizis}, {Giampapa}, {Rutledge},
  {Liebert}, {Mart{\'{\i}}n}, {Basri}, {Fleming}, {Johns-Krull}, {Phan-Bao}, \&
  {Sherry}}]{berger08}
{Berger}, E., {et~al.} 2008, \apj, 673, 1080

\bibitem[{{Bessell} \& {Brett}(1988)}]{bessell88}
{Bessell}, M.~S., \& {Brett}, J.~M. 1988, \pasp, 100, 1134

\bibitem[{{Bochanski} {et~al.}(2010){Bochanski}, {Hawley}, {Covey}, {West},
  {Reid}, {Golimowski}, \& {Ivezi{\'c}}}]{boo10}
{Bochanski}, J.~J., {Hawley}, S.~L., {Covey}, K.~R., {West}, A.~A., {Reid},
  I.~N., {Golimowski}, D.~A., \& {Ivezi{\'c}}, {\v Z}. 2010, \aj, 139, 2679

\bibitem[{{Bochanski} {et~al.}(2011){Bochanski}, {Hawley}, \& {West}}]{boostat}
{Bochanski}, J.~J., {Hawley}, S.~L., \& {West}, A.~A. 2011, \aj, submitted

\bibitem[{{Bochanski} {et~al.}(2007{\natexlab{a}}){Bochanski}, {Munn},
  {Hawley}, {West}, {Covey}, \& {Schneider}}]{boomunn}
{Bochanski}, J.~J., {Munn}, J.~A., {Hawley}, S.~L., {West}, A.~A., {Covey},
  K.~R., \& {Schneider}, D.~P. 2007{\natexlab{a}}, \aj, 134, 2418

\bibitem[{{Bochanski} {et~al.}(2007{\natexlab{b}}){Bochanski}, {West},
  {Hawley}, \& {Covey}}]{bootem}
{Bochanski}, J.~J., {West}, A.~A., {Hawley}, S.~L., \& {Covey}, K.~R.
  2007{\natexlab{b}}, \aj, 133, 531

\bibitem[{{Branham}(2003)}]{branham03}
{Branham}, Jr., R.~L. 2003, \apss, 288, 417

\bibitem[{{Burgasser} {et~al.}(2002){Burgasser}, {Liebert}, {Kirkpatrick}, \&
  {Gizis}}]{burgasser02}
{Burgasser}, A.~J., {Liebert}, J., {Kirkpatrick}, J.~D., \& {Gizis}, J.~E.
  2002, \aj, 123, 2744

\bibitem[{{Cardelli} {et~al.}(1989){Cardelli}, {Clayton}, \& {Mathis}}]{CCM}
{Cardelli}, J.~A., {Clayton}, G.~C., \& {Mathis}, J.~S. 1989, \apj, 345, 245

\bibitem[{{Charbonneau} {et~al.}(2009){Charbonneau}, {Berta}, {Irwin}, {Burke},
  {Nutzman}, {Buchhave}, {Lovis}, {Bonfils}, {Latham}, {Udry}, {Murray-Clay},
  {Holman}, {Falco}, {Winn}, {Queloz}, {Pepe}, {Mayor}, {Delfosse}, \&
  {Forveille}}]{mearth09}
{Charbonneau}, D., {et~al.} 2009, \nat, 462, 891

\bibitem[{{Cincunegui} {et~al.}(2007){Cincunegui}, {D{\'{\i}}az}, \&
  {Mauas}}]{cincunegui07}
{Cincunegui}, C., {D{\'{\i}}az}, R.~F., \& {Mauas}, P.~J.~D. 2007, \aap, 469,
  309

\bibitem[{{Covey} {et~al.}(2007){Covey}, {Ivezi{\'c}}, {Schlegel},
  {Finkbeiner}, {Padmanabhan}, {Lupton}, {Ag{\"u}eros}, {Bochanski}, {Hawley},
  {West}, {Seth}, {Kimball}, {Gogarten}, {Claire}, {Haggard}, {Kaib},
  {Schneider}, \& {Sesar}}]{covey07}
{Covey}, K.~R., {et~al.} 2007, \aj, 134, 2398

\bibitem[{{Covey} {et~al.}(2008{\natexlab{a}}){Covey}, {Ag{\"u}eros}, {Green},
  {Haggard}, {Barkhouse}, {Drake}, {Evans}, {Kashyap}, {Kim}, {Mossman},
  {Pease}, \& {Silverman}}]{champ}
---. 2008{\natexlab{a}}, \apjs, 178, 339

\bibitem[{{Covey} {et~al.}(2008{\natexlab{b}}){Covey}, {Hawley}, {Bochanski},
  {West}, {Reid}, {Golimowski}, {Davenport}, {Henry}, {Uomoto}, \&
  {Holtzman}}]{covey08}
---. 2008{\natexlab{b}}, \aj, 136, 1778

\bibitem[{{Cruz} {et~al.}(2003){Cruz}, {Reid}, {Liebert}, {Kirkpatrick}, \&
  {Lowrance}}]{cruz03}
{Cruz}, K.~L., {Reid}, I.~N., {Liebert}, J., {Kirkpatrick}, J.~D., \&
  {Lowrance}, P.~J. 2003, \aj, 126, 2421

\bibitem[{{Cutri} {et~al.}(2003)}]{2mass}
{Cutri}, R.~M., {et~al.} 2003, {2MASS All Sky Catalog of point sources.}

\bibitem[{{Dahn} {et~al.}(2002)}]{dahn02}
{Dahn}, C.~C., {et~al.} 2002, \aj, 124, 1170

\bibitem[{{Deacon} {et~al.}(2005){Deacon}, {Hambly}, \& {Cooke}}]{deacon05}
{Deacon}, N.~R., {Hambly}, N.~C., \& {Cooke}, J.~A. 2005, \aap, 435, 363

\bibitem[{{Faherty} {et~al.}(2011)}]{faherty11}
{Faherty}, J., {et~al.} 2011, \aj, submitted

\bibitem[{{Fuchs} {et~al.}(2009){Fuchs}, {Dettbarn}, {Rix}, {Beers}, {Bizyaev},
  {Brewington}, {Jahrei{\ss}}, {Klement}, {Malanushenko}, {Malanushenko},
  {Oravetz}, {Pan}, {Simmons}, \& {Snedden}}]{fuchs09}
{Fuchs}, B., {et~al.} 2009, \aj, 137, 4149

\bibitem[{{Fukugita} {et~al.}(1996){Fukugita}, {Ichikawa}, {Gunn}, {Doi},
  {Shimasaku}, \& {Schneider}}]{fukugita96}
{Fukugita}, M., {Ichikawa}, T., {Gunn}, J.~E., {Doi}, M., {Shimasaku}, K., \&
  {Schneider}, D.~P. 1996, \aj, 111, 1748

\bibitem[{{Gizis}(1997)}]{gizis97}
{Gizis}, J.~E. 1997, \aj, 113, 806

\bibitem[{{Gizis} {et~al.}(2000){Gizis}, {Monet}, {Reid}, {Kirkpatrick},
  {Liebert}, \& {Williams}}]{gizis00}
{Gizis}, J.~E., {Monet}, D.~G., {Reid}, I.~N., {Kirkpatrick}, J.~D., {Liebert},
  J., \& {Williams}, R.~J. 2000, \aj, 120, 1085

\bibitem[{{Gliese} \& {Jahrei{\ss}}(1991)}]{gliese91}
{Gliese}, W., \& {Jahrei{\ss}}, H. 1991, {Preliminary Version of the Third
  Catalogue of Nearby Stars}, Tech. rep.

\bibitem[{{Gunn} {et~al.}(1998)}]{gunn98}
{Gunn}, J.~E., {et~al.} 1998, \aj, 116, 3040

\bibitem[{{Gunn} {et~al.}(2006)}]{gunn06}
---. 2006, \aj, 131, 2332

\bibitem[{{Hall}(1996)}]{hall96}
{Hall}, J.~C. 1996, \pasp, 108, 313

\bibitem[{{Hawley} {et~al.}(1996){Hawley}, {Gizis}, \& {Reid}}]{hawley96}
{Hawley}, S.~L., {Gizis}, J.~E., \& {Reid}, I.~N. 1996, \aj, 112, 2799

\bibitem[{{Hawley} \& {Pettersen}(1991)}]{hawley91}
{Hawley}, S.~L., \& {Pettersen}, B.~R. 1991, \apj, 378, 725

\bibitem[{{Hawley} {et~al.}(2002)}]{H02}
{Hawley}, S.~L., {et~al.} 2002, \aj, 123, 3409

\bibitem[{{Hilton} {et~al.}(2010){Hilton}, {West}, {Hawley}, \&
  {Kowalski}}]{hilton10}
{Hilton}, E.~J., {West}, A.~A., {Hawley}, S.~L., \& {Kowalski}, A.~F. 2010,
  ArXiv e-prints

\bibitem[{{Hogg} {et~al.}(2001){Hogg}, {Finkbeiner}, {Schlegel}, \&
  {Gunn}}]{hogg01}
{Hogg}, D.~W., {Finkbeiner}, D.~P., {Schlegel}, D.~J., \& {Gunn}, J.~E. 2001,
  \aj, 122, 2129

\bibitem[{{Ivezi{\'c}} {et~al.}(2004)}]{ivezic04}
{Ivezi{\'c}}, {\v Z}., {et~al.} 2004, Astronomische Nachrichten, 325, 583

\bibitem[{{Ivezic} {et~al.}(2008)}]{lsst}
{Ivezic}, Z., {et~al.} 2008, Serbian Astronomical Journal, 176, 1

\bibitem[{{Ivezi{\'c}} {et~al.}(2008)}]{ivezic08}
{Ivezi{\'c}}, {\v Z}., {et~al.} 2008, \apj, 684, 287

\bibitem[{{Johnson} \& {Apps}(2009)}]{johnson09}
{Johnson}, J.~A., \& {Apps}, K. 2009, \apj, 699, 933

\bibitem[{{Jones} {et~al.}(2011){Jones}, {West}, \& {Foster}}]{jones11}
{Jones}, D., {West}, A.~A., \& {Foster}, J. 2011, \aj, submitted

\bibitem[{{Juri{\'c}} {et~al.}(2008)}]{juric08}
{Juri{\'c}}, M., {et~al.} 2008, \apj, 673, 864

\bibitem[{{Kowalski} {et~al.}(2009){Kowalski}, {Hawley}, {Hilton}, {Becker},
  {West}, {Bochanski}, \& {Sesar}}]{kowalski09}
{Kowalski}, A.~F., {Hawley}, S.~L., {Hilton}, E.~J., {Becker}, A.~C., {West},
  A.~A., {Bochanski}, J.~J., \& {Sesar}, B. 2009, \aj, 138, 633

\bibitem[{{Kruse} {et~al.}(2010){Kruse}, {Berger}, {Knapp}, {Laskar}, {Gunn},
  {Loomis}, {Lupton}, \& {Schlegel}}]{kruse10}
{Kruse}, E.~A., {Berger}, E., {Knapp}, G.~R., {Laskar}, T., {Gunn}, J.~E.,
  {Loomis}, C.~P., {Lupton}, R.~H., \& {Schlegel}, D.~J. 2010, \apj, 722, 1352

\bibitem[{{Leggett} {et~al.}(1996){Leggett}, {Allard}, {Berriman}, {Dahn}, \&
  {Hauschildt}}]{leggett96}
{Leggett}, S.~K., {Allard}, F., {Berriman}, G., {Dahn}, C.~C., \& {Hauschildt},
  P.~H. 1996, \apjs, 104, 117

\bibitem[{{Leggett} {et~al.}(2001){Leggett}, {Allard}, {Geballe}, {Hauschildt},
  \& {Schweitzer}}]{leggett01}
{Leggett}, S.~K., {Allard}, F., {Geballe}, T.~R., {Hauschildt}, P.~H., \&
  {Schweitzer}, A. 2001, \apj, 548, 908

\bibitem[{{L{\'e}pine} {et~al.}(2007){L{\'e}pine}, {Rich}, \&
  {Shara}}]{lepine07}
{L{\'e}pine}, S., {Rich}, R.~M., \& {Shara}, M.~M. 2007, \apj, 669, 1235

\bibitem[{{L{\'e}pine} \& {Scholz}(2008)}]{lepine08}
{L{\'e}pine}, S., \& {Scholz}, R. 2008, \apjl, 681, L33

\bibitem[{{L{\'e}pine} \& {Shara}(2005)}]{lspm}
{L{\'e}pine}, S., \& {Shara}, M.~M. 2005, \aj, 129, 1483

\bibitem[{{Luyten}(1979)}]{nltt}
{Luyten}, W.~J. 1979, {LHS catalogue. A catalogue of stars with proper motions
  exceeding 0''5 annually}, ed. {Luyten, W.~J.}

\bibitem[{{Munn} {et~al.}(2004){Munn}, {Monet}, {Levine}, {Canzian}, {Pier},
  {Harris}, {Lupton}, {Ivezi{\'c}}, {Hindsley}, {Hennessy}, {Schneider}, \&
  {Brinkmann}}]{munn04}
{Munn}, J.~A., {et~al.} 2004, \aj, 127, 3034

\bibitem[{{Munn} {et~al.}(2008){Munn}, {Monet}, {Levine}, {Canzian}, {Pier},
  {Harris}, {Lupton}, {Ivezi{\'c}}, {Hindsley}, {Hennessy}, {Schneider}, \&
  {Brinkmann}}]{munn08}
---. 2008, \aj, 136, 895

\bibitem[{{Nordstr{\"o}m} {et~al.}(2004){Nordstr{\"o}m}, {Mayor}, {Andersen},
  {Holmberg}, {Pont}, {J{\o}rgensen}, {Olsen}, {Udry}, \&
  {Mowlavi}}]{nordstrom04}
{Nordstr{\"o}m}, B., {et~al.} 2004, \aap, 418, 989

\bibitem[{{Pier} {et~al.}(2003){Pier}, {Munn}, {Hindsley}, {Hennessy}, {Kent},
  {Lupton}, \& {Ivezi{\'c}}}]{pier03}
{Pier}, J.~R., {Munn}, J.~A., {Hindsley}, R.~B., {Hennessy}, G.~S., {Kent},
  S.~M., {Lupton}, R.~H., \& {Ivezi{\'c}}, {\v Z}. 2003, \aj, 125, 1559

\bibitem[{{Pineda} {et~al.}(2011){Pineda}, {West}, {Bochanski}, \&
  {Burgaser}}]{pineda10}
{Pineda}, J.~S., {West}, A.~A., {Bochanski}, J.~J., \& {Burgaser}, A.~J. 2011,
  \aj, submitted

\bibitem[{{Rauscher} \& {Marcy}(2006)}]{rm06}
{Rauscher}, E., \& {Marcy}, G.~W. 2006, \pasp, 118, 617

\bibitem[{{Reid} {et~al.}(1995{\natexlab{a}}){Reid}, {Hawley}, \&
  {Gizis}}]{pmsu1}
{Reid}, I.~N., {Hawley}, S.~L., \& {Gizis}, J.~E. 1995{\natexlab{a}}, \aj, 110,
  1838

\bibitem[{{Reid} {et~al.}(1995{\natexlab{b}}){Reid}, {Hawley}, \&
  {Mateo}}]{reid95}
{Reid}, N., {Hawley}, S.~L., \& {Mateo}, M. 1995{\natexlab{b}}, \mnras, 272,
  828

\bibitem[{{Rojas-Ayala} {et~al.}(2010){Rojas-Ayala}, {Covey}, {Muirhead}, \&
  {Lloyd}}]{babs10}
{Rojas-Ayala}, B., {Covey}, K.~R., {Muirhead}, P.~S., \& {Lloyd}, J.~P. 2010,
  \apjl, 720, L113

\bibitem[{{Salim} \& {Gould}(2003)}]{salim03}
{Salim}, S., \& {Gould}, A. 2003, \apj, 582, 1011

\bibitem[{{Schlegel} {et~al.}(1998){Schlegel}, {Finkbeiner}, \& {Davis}}]{sfd}
{Schlegel}, D.~J., {Finkbeiner}, D.~P., \& {Davis}, M. 1998, \apj, 500, 525

\bibitem[{{Schmidt} {et~al.}(2007){Schmidt}, {Cruz}, {Bongiorno}, {Liebert}, \&
  {Reid}}]{schmidt07}
{Schmidt}, S.~J., {Cruz}, K.~L., {Bongiorno}, B.~J., {Liebert}, J., \& {Reid},
  I.~N. 2007, \aj, 133, 2258

\bibitem[{{Schmidt} {et~al.}(2010{\natexlab{a}}){Schmidt}, {West}, {Burgasser},
  {Bochanski}, \& {Hawley}}]{schmidt_blue}
{Schmidt}, S.~J., {West}, A.~A., {Burgasser}, A.~J., {Bochanski}, J.~J., \&
  {Hawley}, S.~L. 2010{\natexlab{a}}, \aj, 139, 1045

\bibitem[{{Schmidt} {et~al.}(2010{\natexlab{b}}){Schmidt}, {West}, {Hawley}, \&
  {Pineda}}]{schmidt_sam}
{Schmidt}, S.~J., {West}, A.~A., {Hawley}, S.~L., \& {Pineda}, J.~S.
  2010{\natexlab{b}}, \aj, 139, 1808

\bibitem[{{Segura} {et~al.}(2010){Segura}, {Walkowicz}, {Meadows}, {Kasting},
  \& {Hawley}}]{segura10}
{Segura}, A., {Walkowicz}, L., {Meadows}, V., {Kasting}, J., \& {Hawley}, S.
  2010, ArXiv e-prints

\bibitem[{{Smith} {et~al.}(2002)}]{smith02}
{Smith}, J.~A., {et~al.} 2002, \aj, 123, 2121

\bibitem[{{Smol{\v c}i{\'c}} {et~al.}(2004){Smol{\v c}i{\'c}}, {Ivezi{\'c}},
  {Knapp}, {Lupton}, {Pavlovski}, {Iliji{\'c}}, {Schlegel}, {Smith}, {McGehee},
  {Silvestri}, {Hawley}, {Rockosi}, {Gunn}, {Strauss}, {Fan}, {Eisenstein}, \&
  {Harris}}]{Smolcic04}
{Smol{\v c}i{\'c}}, V., {et~al.} 2004, \apjl, 615, L141

\bibitem[{{Tucker} {et~al.}(2006)}]{tucker06}
{Tucker}, D.~L., {et~al.} 2006, Astronomische Nachrichten, 327, 821

\bibitem[{{Walkowicz} \& {Hawley}(2009)}]{walkowicz09}
{Walkowicz}, L.~M., \& {Hawley}, S.~L. 2009, \aj, 137, 3297

\bibitem[{{Walkowicz} {et~al.}(2004){Walkowicz}, {Hawley}, \&
  {West}}]{walkowicz04}
{Walkowicz}, L.~M., {Hawley}, S.~L., \& {West}, A.~A. 2004, \pasp, 116, 1105

\bibitem[{{West} \& {Basri}(2009)}]{westbasri09}
{West}, A.~A., \& {Basri}, G. 2009, \apj, 693, 1283

\bibitem[{{West} {et~al.}(2006){West}, {Bochanski}, {Hawley}, {Cruz}, {Covey},
  {Silvestri}, {Reid}, \& {Liebert}}]{west06}
{West}, A.~A., {Bochanski}, J.~J., {Hawley}, S.~L., {Cruz}, K.~L., {Covey},
  K.~R., {Silvestri}, N.~M., {Reid}, I.~N., \& {Liebert}, J. 2006, \aj, 132,
  2507

\bibitem[{{West} \& {Hawley}(2008)}]{westhawley08}
{West}, A.~A., \& {Hawley}, S.~L. 2008, \pasp, 120, 1161

\bibitem[{{West} {et~al.}(2008){West}, {Hawley}, {Bochanski}, {Covey}, {Reid},
  {Dhital}, {Hilton}, \& {Masuda}}]{west08}
{West}, A.~A., {Hawley}, S.~L., {Bochanski}, J.~J., {Covey}, K.~R., {Reid},
  I.~N., {Dhital}, S., {Hilton}, E.~J., \& {Masuda}, M. 2008, \aj, 135, 785

\bibitem[{{West} {et~al.}(2005){West}, {Walkowicz}, \& {Hawley}}]{west05}
{West}, A.~A., {Walkowicz}, L.~M., \& {Hawley}, S.~L. 2005, \pasp, 117, 706

\bibitem[{{West} {et~al.}(2004){West}, {Hawley}, {Walkowicz}, {Covey},
  {Silvestri}, {Raymond}, {Harris}, {Munn}, {McGehee}, {Ivezi{\'c}}, \&
  {Brinkmann}}]{west04}
{West}, A.~A., {et~al.} 2004, \aj, 128, 426

\bibitem[{{Woolf} {et~al.}(2009){Woolf}, {L{\'e}pine}, \&
  {Wallerstein}}]{woolf09}
{Woolf}, V.~M., {L{\'e}pine}, S., \& {Wallerstein}, G. 2009, \pasp, 121, 117

\bibitem[{{Yanny} {et~al.}(2009)}]{segue}
{Yanny}, B., {et~al.} 2009, \aj, 137, 4377

\bibitem[{{York} {et~al.}(2000)}]{york00}
{York}, D.~G., {et~al.} 2000, \aj, 120, 1579

\end{thebibliography}


\end{document}